\documentclass[pdflatex,sn-mathphys-num]{sn-jnl}% Math and Physical Sciences Numbered Reference Style
\usepackage{graphicx}%
\usepackage{multirow}%
\usepackage{amsmath,amssymb,amsfonts}%
\usepackage{amsthm}%
\usepackage{mathrsfs}%
\usepackage[title]{appendix}%
\usepackage{xcolor}%
\usepackage{textcomp}%
\usepackage{manyfoot}%
\usepackage{booktabs}%
\usepackage{algorithm}%
\usepackage{algorithmicx}%
\usepackage{algpseudocode}%
\usepackage{listings}%
\usepackage{subcaption}
\theoremstyle{thmstyleone}%
\theoremstyle{thmstyletwo}%

\theoremstyle{thmstylethree}%

\begin{document}

% \title[Article Title]{Article Title}

% %%=============================================================%%
% %% GivenName	-> \fnm{Joergen W.}
% %% Particle	-> \spfx{van der} -> surname prefix
% %% FamilyName	-> \sur{Ploeg}
% %% Suffix	-> \sfx{IV}
% %% \author*[1,2]{\fnm{Joergen W.} \spfx{van der} \sur{Ploeg} 
% %%  \sfx{IV}}\email{iauthor@gmail.com}
% %%=============================================================%%

% \author*[1,2]{\fnm{First} \sur{Author}}\email{iauthor@gmail.com}



% \affil*[1]{\orgdiv{Department}, \orgname{Organization}, \orgaddress{\street{Street}, \city{City}, \postcode{100190}, \state{State}, \country{Country}}}

% \affil[2]{\orgdiv{Department}, \orgname{Organization}, \orgaddress{\street{Street}, \city{City}, \postcode{10587}, \state{State}, \country{Country}}}

% \affil[3]{\orgdiv{Department}, \orgname{Organization}, \orgaddress{\street{Street}, \city{City}, \postcode{610101}, \state{State}, \country{Country}}}

\title{Chaotic Oscillator Associative Memory}

% Use letters for affiliations, numbers to show equal authorship (if applicable) and to indicate the corresponding author
\author[a,b]{Nurani Rajagopal Rohan}
\author[a,c]{V. Srinivasa Chakravarthy}
\author*[b,c]{Sayan Gupta}\email{sayan@iitm.ac.in}

\affil[a]{Laboratory of Computational Neuroscience, Department of Biotechnology, Indian Institute of Technology Madras 600036 India
}
\affil[b]{The Uncertainty Lab, Department of Applied Mechanics, Indian Institute of Technology Madras 600036 India}
\affil[c]{Center for Complex Systems \& Dynamics IIT Madras 600036 India}


\abstract{Associative memory models retrieve stored information through content-based addressing, mimicking the neural processes of animal brains. The classical Hopfield network-based models store memories as vectors of discrete values and have good storage capacity but do not consider the role of neuronal synchronization in memory storage and retrieval as observed in brains. This is addressed in phase-oscillator-based models which store memories as  time-dependent phase-synchronized states, but suffer from instability and low capacity. The present study addresses these challenges through a novel chaotic oscillator-based associative memory model, by defining a phase relationship in chaotic systems and encoding memory as synchronized states of these phases.  The underlying chaos in the network is shown to significantly improve both storage and retrieval  and offer insights into the dynamics of memory retrieval.}




\maketitle

\section{Introduction}
\label{introduction}
Associative memory models are a powerful class of machine learning tools designed for pattern recognition, error correction and learning. Inspired by the mechanisms of the animal brain, these models differ from  traditional von Neumann computing  architecture, where data retrieval relies on explicit indexing. Instead, associative memory systems retrieve data based on content similarity, mimicking the brain’s ability to recall memories from partial or noisy information through association, and consequently have additional error correcting feature which are otherwise handled using software. As the stored information itself functions as the retrieval key, these are also referred to as content-addressable systems \cite{hopfieldS, hopfieldC}

In associative memory models, memories refer to the stored content (or patterns). The seminal Hopfield associative memory model \cite{hopfieldS} is a fully connected recurrent neural network, where each neuron takes binary states ($\pm 1$) and the weights of the connections are  based on  Hebbian learning \cite{Hebb}. The Hopfield network minimizes an energy function, with stored patterns (or data) corresponding to  local minima in the energy landscape, representing stable states. When a partial or a noisy version of a stored pattern is presented on which the network has been trained, the neurons update asynchronously based on the weighted sum of the inputs and the system iteratively converges to the closest stable state in a process that is referred to as retrieval. From a dynamical systems perspective,  memories in the Hopfield network function as point attractors in a multi-dimensional state space, where  noisy inputs evolve towards the attractor  within whose basins they lie. This concept paved the way for the broader class of neural associative memory models \cite{hopfieldS, amit, amarinet}. With increasing  number of attractors stored in the network and the corresponding intertwining of their basins, even small perturbations lead to retrieval failure, implying loss of stability for all the patterns. 

For a $N$-neuron Hopfield network, the critical storage capacity $\alpha_c$, defined as the maximum number of memories that can be stored without loss of stability, has been shown to be $\alpha_c=0.138N$. This means that a Hopfield network with 100 neurons can reliably store and retrieve fewer than 14 patterns.  Several extensions of the Hopfield network include those with additional layers, different activation functions, learning algorithms and complex update rules. These dense associative memory models though enhance storage capacity and model more complex memory structures\cite{JCook_1989},  are very different from the traditional Hopfield model and their training can be cumbersome. Alternatives also include generalization of the binary neurons to multi-states - either discrete or continuous, but in general these have been found to lead to lower storage capacity.

Despite the significant advancements in Hopfield network memory models, they are inherently designed for storing and retrieving static patterns, making them fundamentally unsuitable for handling dynamic or time-dependent sequences essential to real-world cognitive tasks  such as speech recognition, motor control, and dynamic sensory processing.  In contrast, biological memory operates in a highly dynamic environment, where information processing and recall are influenced by time-dependent signals. Neuroscientific evidence suggests that cognitive processes, particularly memory formation, consolidation, and retrieval are closely linked to  neural oscillations in  the theta (4–8 Hz) and gamma (30–80 Hz) frequency bands  \cite{fell, duzel, kahana2006cognitive, KLIMESCH200831}. Research has shown that spatial and episodic memories are encoded through synchronized oscillatory activity in the brain. For instance, studies on hippocampal cells indicate that memories related to spatial locations \cite{keefe} and objects \cite{siegel} are stored as phase relationships between neuronal oscillations in the theta band.

Motivated by these observations, researchers have turned to biologically inspired models such as oscillatory neural networks \cite{abbott1990network}, which mimic this phase-based encoding by storing binary patterns as phase relationships among coupled oscillators. Unlike Hopfield networks where memory states are point attractors, these models  encode memories as limit cycles or phase locked attractors emerging from synchronized phase   dynamics. 
In an oscillator based memory network, each memory unit is typically modeled as a phase oscillator, often using the  Kuramoto model \cite{acebron2005kuramoto}
\begin{equation}
\label{eq:general_oscillator_associative_memory}
    \dot{\theta_i} = \omega_i +  \sum_{j=1}^N\mathfrak{F}_{ij}(\theta_i, \theta_j),\,\,\, i=1, \cdots, N,
\end{equation}
where, $\omega_i$ is the intrinsic frequency of an oscillator and $\mathfrak{F}_{ij}$ is a $2\pi$-periodic function that models the pair-wise coupling between two oscillators. A stored pattern corresponds to a stable phase-locked state, and upon receiving a partial or noisy input, the system dynamically evolves to the nearest attractor.  When the coupling uses only the first Fourier mode, $\mathfrak{F}_{ij} \propto \sin(\theta_i - \theta_j)$, studies show that regardless of network size, error-free retrieval becomes unstable once the number of memories $p$ stored are more than 2 \cite{aonishi}.

When the number of stored patterns exceeds two ($p>2$), retrieval quality drops significantly — achieving only about $69\%$ overlap with the correct pattern.  This reflects a convergence to mixture states that overlap with multiple stored patterns, rather than a clean retrieval and can be attributed to the first Fourier mode in the coupling, which models only symmetric interactions. Including second and higher order Fourier modes in the coupling model enables the interactions to be more nonlinear and selective, allowing richer dynamics and less interference between stored patterns, thereby enhancing the storage capacity \cite{nishikawa,follman}.

Oscillatory models offer significant advantages for neuromorphic computing and hardware implementation, as they can be physically realized using resonant systems such as lasers \cite{hoppensteadt} or MEMS resonators \cite{hoppensteadt_mem}. These implementations promise  energy efficiency, high-speed memory access, and greater scalability compared to traditional Hopfield networks. However, despite their biological plausibility and hardware potential, oscillatory memory models face several critical challenges. A major limitation is their low storage capacity relative to Hopfield networks, due to  the limited number of stable phase-locked states. To mitigate these issues, a recent hybrid approach \cite{delacour2021mapping, todri2024computing}  integrates Hopfield-like energy minimization with oscillatory dynamics via phase-locked loops, optimizing phase behavior to enhance both stability and storage. Additionally, neuromorphic computing and Field 
 Programable Gate Arrays (FPGA) have provided innovative solutions for efficient phase-computing in oscillatory associative memory (OAM) models \cite{nunez2021oscillatory, todri2021frequency, abernot2021digital}. % balance stability and capacity, improving retrieval performance. 
However, significant opportunities remain for further advancement. 

The present paper proposes a novel associative memory model based on a network of chaotic oscillators, the idea being inspired by the seminal work on olfactory processing by Freeman and colleagues \cite{skf, freeman1972waves, freeman_eeg_spatial, freeman1987simulation, freeman1979eeg}. Their electrophysiological experiments on the rabbit’s olfactory bulb revealed that the spatio-temporal activity induced by the odor did not converge to fixed points or limit cycles but instead exhibited low-dimensional chaotic dynamics. Notably, even familiar odors produced amplitude-modulated chaotic patterns that varied across trials \cite{skf}. Freeman hypothesized that this variability is functional rather than noise, and that such chaotic activity supports memory retrieval by preventing entrapment in cyclic dynamics, thereby maintaining access to all learned stimuli.  

Building on these insights, several studies have explored the role of chaos in enhancing storage and retrieval in associative memory models \cite{csnn_nakagawa, camm_nakagawa, adachi1997, aihara1990, calitoiu2007, HE20082794, Kushibe_amstcc, 726629, tsuda1992dynamic}. One notable example is the Adachi chaotic neural network (AdNN) \cite{adachi1997, aihara1990, calitoiu2007}, which is implemented using a discrete four-variable system with two or three internal states. ADNN modifies the original Hopfield model by incorporating the Ikeda map and delayed feedback to model Freeman’s observations. Given a stimulus, the network demonstrates long-term transient dynamics and  approaches  the stored pattern chaotically, without attaining long term convergence. The model is not designed to enhance  storage capacity, but is useful in locating optimal extrema in optimization problems. 
Unlike existing chaotic models of associative memory that relies on the original Hopfield architecture, this paper proposes a novel Chaotic Oscillator Associative Memory (COAM) comprising of a network of chaotic oscillators, with the coupling strengths being trained through Hebbian learning. The key distinction from existing OAM models lies in the higher state space dimensionality of each neuron within the network, with the challenge lying  in establishing the phase relationships between them - an essential principle underlying phase-based encoding in OAM architectures. Hardware implementation of chaotic systems, whether analog \cite{analog_chaotic_implementation_1, analog_chaotic_implementation_2} or digital \cite{estudillo2023fpga, chaotic_FPGA}, is extensively studied; COAM as a memory system can be implemented in such devices. 

\section{The Memory Model}
\label{Sec:CAM}
%This paper proposes a novel generalization of OAMs by building an associative memory model comprising of a network of coupled chaotic systems. 
Unlike traditional OAMs, where each fundamental neuron is represented solely by its phase (see Eq.(\ref{eq:general_oscillator_associative_memory})), %comprising of only the phase as the state variable, 
chaotic oscillators operate in a state space of at least three dimensions. To utilize the principle of memory encoding through phase relationships in such systems, the  challenge lies in formulating a meaningful  description of phase from the higher-dimensional chaotic dynamics and which is also consistent with stable, phase-locked synchronization states as an emergent phenomena across the network.
%capturing emergent phenomenon such as  stable, phase-locked synchronization states across the network.
%As memory encoding in OAMs is through phase synchronization of stable phase locked solutions, a key question that first needs to be addressed is on developing a phase description in the chaotic dynamics.
As the topological features of chaotic attractors lack generality, a single unequivocal definition of phase for chaotic systems is difficult.
%cannot be given, leading to the development of various definitions that incorporate the topological features of the chaotic attractor. 
However, an explicit form for the phase based on the general concept of curvature for an arbitrary curve is possible. %For a two-dimensional path represented by $\vec{r} = (u,v)$, 
For example, if the projection of a chaotic trajectory on some plane is represented as a two-dimensional vector $\vec{r} = (u,v)$ and this path has a positive curvature, defined as $R = (\dot{u}^2 + \dot{v}^2)^{3/2}/\left|\dot{v} \ddot{u} - \dot{u} \ddot{v}\right|$, and if the variable
 $\phi = \arctan(\dot{v}/\dot{u})$ increases monotonically in time, then $\phi$  can be considered as a phase of the system \cite{Kurths_strong_chaotic_systems}. %The definitions of phase and frequency apply to any dynamical system if the projection of the phase trajectory on some plane is a curve with a positive curvature and have been shown to work for a wide range of chaotic oscillators \cite{Kurths_strong_chaotic_systems}. In particular for the R\"ossler system, projecting the system on $(\dot{x}, \dot{y})$, the chaotic trajectories rotate around the origin and the phase can be defined as, $\phi = \arctan(\dot{y}/\dot{x}).$ 
 This underlying principle of this definition is based on projecting higher dimension dynamics onto a plane and hence the phase description is an approximation. Moreover, as the system dynamics is chaotic, this phase representation will not be periodic. The deviation of phase of a chaotic system from that of a periodic oscillator is measured by the diffusion constant $D_p$ \cite{Kurths_strong_chaotic_systems},  defined as $\langle (\phi(t) - \langle \phi(t) \rangle)^2\rangle = 2D_p\, t$, where the operator $\langle \cdot\rangle$ represents temporal averaging. %For a periodic oscillator, $D_p$ is zero, whereas for a chaotic system, $D_p$ is nonzero. 
 For chaotic systems with small $D_p$, the phase can be treated as a piecewise continuous function (that increases linearly in time) of successive intersections of the chaotic trajectory on a suitably chosen Poincar\'e surface. 
   %simpler phase definition based on a Poincar\'e surface can be used. In this case, the phase can be treated as a piecewise continuous function that increases linearly with time between successive intersections of a suitably chosen Poincaré surface. 
 The phase dynamics on the plane depends on its instantaneous frequency and its amplitude,  %of the latter and therefore its phase has  been shown to typically vary with amplitude  
 and can be qualitatively written as \cite{spectral_broadening, PhysRevLett.76.1804, Pikovsky2003} \begin{align}\label{eq:phase_of_a_general_chaotic_oscillator}
  %  A_{n+1} &= M(A_n),\nonumber\\
   {\dot {\phi}} &= \omega(A) \equiv \overline{\omega} + F(A),
\end{align}
where, $\overline{\omega}$ is the mean frequency of the system, $A$ is the amplitude of the chaotic oscillations and $F(A)$ is the effect of the amplitude on the frequency and is broad-banded as the dynamics is chaotic; see \cite{Pikovsky2003} for more details.  Eq.(\ref{eq:phase_of_a_general_chaotic_oscillator}) is qualitatively similar to the phase of a periodic oscillator subjected to external noise, where $\overline{\omega}$ is its angular frequency and $F(A)$ can be treated as the effective noise. The power spectrum of the state variables of these class of chaotic oscillators are typically characterized by a well defined peak further reinforcing this viewpoint. On the other hand, the power spectrum exhibits significant broader bandwidth for chaotic oscillators with high $D_p$, and the absence of a single characteristic time scale. Consequently, the analogy between chaotic phase dynamics and that of a periodic oscillator perturbed by external noise cannot be  applied to these oscillators.

This phase representation of chaotic oscillators (with small $D_p \approx \mathcal{O}(10^{-3})$) provides a framework for a generalized model for COAM, comprising of a weighted, fully connected network of $N$ chaotic oscillators.  The equations of motion of each individual COAM neuron can be expressed as
%for describing the phase dynamics in a network of chaotic oscillators through the equation 
\begin{equation}
\dot{\phi}_i = \omega_i + F_i(A_i) + \epsilon \sum_{j=1}^N~ \mathfrak{F}_{ij}(\phi_i, \phi_j), \,\,\, i=1, \hdots, N,
\label{eq:coam1}
\end{equation}
where, $\omega_i =\bar{\omega}_i$ represents the intrinsic averaged frequency of the $i$-th chaotic oscillator, $F_i(A_i)$ is the influence of the oscillator's amplitude $A_i$ on its phase evolution, $\epsilon$ is the coupling strength, and $\mathfrak{F}_{ij}$ is a $2\pi$-periodic function that models the pairwise interaction between the chaotic oscillators in the network. Eq.(\ref{eq:coam1}) can be viewed as a generalization of the OAM model in Eq.(\ref{eq:general_oscillator_associative_memory}). The phase coupling between the oscillators is taken to be bidirectional, and following the developments in OAMs,  is assumed to comprise of higher order Fourier modes, such that
\begin{align}\label{eq:qualitative_CAM}
     \epsilon \sum_{j=1}^N~ \mathfrak{F}_{ij}(\phi_i, \phi_j) =  \epsilon \sum_{j=1}^N  C(t)~ J_{ij}\sin(\phi_{j} - \phi_{i} - \Theta_{ij}) \nonumber\\
    \qquad + \sum_{k = 2}^{K}\left(\frac{\epsilon_k}{N} \sum_{j=1}^N 
    ~\sin~k(\phi_j - \phi_i)\right),
\end{align}
where, $\epsilon J_{ij}$ is the coupling strength,   $\Theta_{ij}$ is the synaptic delay,   $K \in \mathbb{N}$ is the number of higher order Fourier modes in the pairwise coupling,  $\epsilon_k$ are their corresponding strengths, and $C(t)$ is a function of amplitudes $A_i$. 

As in OAM models, memory patterns are encoded into the network as  synchronized phase states $\{\phi_i\}_{i=1}^N$, and are represented  as  complex vectors. The strength of the primary Fourier mode connections is based on Hebbian learning of $p$ stored patterns, and is given by
\begin{equation}\label{eq:phase_weight}
    {\bf W}_{ij} = \frac{1}{N}\sum_{\eta=1}^p \mathfrak{b}^{(\eta)}_i \overline{\mathfrak{b}}^{(\eta)}_j = J_{ij} \exp{(\mathrm{i}\Theta_{ij})}.
\end{equation}
Here,  the $\eta$-th stored pattern  is encoded in terms of the   phase of the oscillators $\{\theta_i^{(\eta)}\}_{i=1} ^N$ and are represented as a complex vector $ \mathfrak{ b}^{(\eta)} = (\mathfrak{b}_1^{(\eta)} \dots \mathfrak{b}^{(\eta)}_N)^\mathrm{T}$,  %each 
where $\mathfrak{b}_i^{(\eta)} = \exp(\mathrm{i}\theta_i^{(\eta)})$, %$\eta = 1 \cdots p$, 
$(\overline{\cdot})$ represents the complex conjugation and i = $\sqrt{-1}$. The loading parameter is defined as $\alpha = p/N$.
Phase synchronization in OAMs implies that the time-dependent phase differences ($\phi_i - \phi_j$) converge asymptotically to constant values. However, due to  $F_i(A_i)$, the convergence of the phase differences in COAM is expected to be accompanied by persistent noisy fluctuations. Nevertheless, the time-averaged phase differences are effectively constant, rendering solutions with similar averaged phase differences to be qualitatively equivalent. When the stored patterns are binary, the network undergoes anti-phase synchronization and phase differences form two clusters, given by
\begin{eqnarray}
    \vert \phi_i - \phi_j \vert  \approx \left\{
	\begin{array}{ll}
		0,  & \mbox{if } \mathfrak{b}_i = \mathfrak{b}_j , \\
		\pi, & \mbox{if } \mathfrak{b}_i \not =  \mathfrak{b}_j .
	\end{array}
\right.
\label{eq:COAM2}
\end{eqnarray}
In such a case, the phase variable in Eq.(\ref{eq:qualitative_CAM}) has $2^N$ fixed points  corresponding to all binary patterns of length $N$. The memory retrieval quality is defined in terms of the measure
\begin{equation}\label{eq:continuous_overlap_m}
    m = \frac{1}{N}\sum_{i=1}^N \overline{\mathfrak{b}}^{(\eta)}_i s_i.
\end{equation}
that quantifies the overlap between the system state vector given by $s_i = \exp(\mathrm{i}\theta_i)$, $i=1,\hdots, N$ and any pattern $\mathfrak{b}^{(\eta)}$.

\subsection*{COAM : R\"ossler system}
%A model for the COAM comprising of a network of $N$ weakly coupled R\"ossler systems, arranged in a weighted, fully connected topology is given. 
The traditional R\"{o}ssler system is expressed in terms of  $x_i$, $y_i$ and $z_i$ as the state variables. However, as the information of the phase is central to the idea of memory encoding, it is more convenient to express the equations  in cylindrical polar coordinates $(A_i,\phi_i,z_i)$, where,  phase $\phi_i=\arctan{y_i}/{x_i}$ and amplitude $A_i=(x_i^2 + y_i^2)^{1/2}$ are the state variables apart from $z_i$. Each R\"ossler neuron is bidirectionally coupled to every other unit through both the $A_i$ and $\phi_i$ state variables. 
%Following the developments in OAMs, the coupling in the phase variable is considered to have higher order Fourier modes. 
The equations of motion of each individual COAM neuron can therefore be expressed as in Eq.(\ref{eq:CAM_with_second_order_coupling}),

\begin{align}\label{eq:CAM_with_second_order_coupling}
    \dot{A}_{i} &= \mathfrak{a}~ A_{i}~ \sin^2 ~\phi_{i} - z_{i}~ \cos~ \phi_{i} + \epsilon A_{i}\sum_{i=1}^N {J_{ij}} \cos(\phi_{j} - \phi_{i} -\Theta_{ij}), \nonumber \\
    \dot{\phi}_{i} &= \omega_{i} + \mathfrak{a}~ \sin~ \phi_{i} ~\cos ~\phi_{i} + z_{i} / A_{i} ~\sin~ \phi_{i} +\\ & \epsilon\frac{1}{A_{i}}\sum_{i=1}^NJ_{ij}A_{j}\sin(\phi_{j} - \phi_{i} - \Theta_{ij}) + \sum_{k = 2}^{K}\left(\frac{\epsilon_k}{N} \sum_{j=1}^N 
    ~\sin~k(\phi_j - \phi_i)\right),\nonumber \\
    \dot{z}_{i} &= \mathfrak{f} - \mathfrak{c}~z_{i} + A_{i} ~ z_{i} ~ \cos ~\phi_{i},\nonumber 
\end{align}

where, $a$, $\mathfrak{f}$, and $c$ are the Rössler parameters. The rest of the parameters remain the same as previously defined in Eq.(\ref{eq:qualitative_CAM}). 
% Particularly for Eq.(\ref{eq:CAM_with_second_order_coupling}, the $\phi$ equation contains the term $\mathfrak{a}\sin 2\phi$ which acts as a subharmonic injection locking term and has been shown to promote antiphase synchronization in networks of coupled oscillators \cite{wang2021solving}. However, 
 %The model demonstrates chaotic behavior for all $\mathfrak{c} > 8.7$. 
%Additionally, it aligns with biological observations, such as the olfactory bulb dynamics in deep anesthesia states described by Skarda and Freeman \cite{skf}, which can be replicated by appropriately choosing R\"ossler parameters $\mathfrak{(a, b, c)}$, leading to point attractor dynamics. 
%However, a limitation of the model in its current form is its inability to exhibit long-term transient dynamics corresponding to an "unknown" state when exposed to novel patterns, as described by Adachi et al. and Ke and Oomen \cite{adachi1997, LNN}. }

% \section{Discussion}
% %%\label{}
% \lipsum[4]

% \section{Summary and conclusions}
% %%\label{}
% \lipsum[1-4]

\section{Results and Discussions}
\label{Sec:Results}
The performance of COAM in encoding patterns in terms of the phase and their retrieval capabilities are investigated through a numerical example comprising of a network of R\"ossler systems. The strength of the couplings  are determined via Hebbian learning of the memories being stored in the network. The network is stimulated with a partial memory and the retrieval dynamics is quantified in terms of an overlap measure  $m$. The analysis is restricted to the case where patterns are binary. More details about the specifics of the model and the algorithm is presented later in the Methods section and  in the appendix.

\subsection{Phase Synchronization in  Chaotic Oscillators}
The concept of phase in chaotic systems is first examined in terms of the synchronization of the phases in two coupled  R\"ossler systems, with only first Fourier mode coupling (Eq.(\ref{eq:CAM_with_second_order_coupling}) with $N=2$, $K=1$). The coupling topology strengths are encoded through the weight matrix $\bf W$ (Eq.(\ref{eq:phase_weight}) with $p=1$).
%The first question that needs to be addressed is what constitutes phase synchronization in  a network of chaotic oscillators. As a first step, consider the case of $N=2$ and $K=1$ of the system defined in Eq.(\ref{eq:CAM_with_second_order_coupling}). This is essentially  two coupled R\"ossler system with only first order Fourier mode in the coupling. The  weight matrix $\bf W$ is defined for $p=1$. 
The difference in phase between the two R\"ossler systems,  $\phi_1-\phi_2$, is a  function of time and on achieving phase   synchronization is expected to asymptotically approach a constant  $\Theta_{12}$. This result can be shown analytically for the special case when $\omega_1=\omega_2 =\omega$, and when  the amplitude $A_{1,2}$ varies slowly over rotation of the phases. %The corresponding system can be analyzed by averaging over these  phase variables $\phi_{1,2}$. 
Introducing ``slow'' phases $\vartheta_{1,2}$ according to $\phi_{1,2} = \omega t + \vartheta_{1,2}$, and time period averaging the equation corresponding to $\phi_{1,2}$, one can get an equation in $\vartheta_{1,2}$, given by %$\dot{\vartheta_1} - \dot{\vartheta_2} =$
\begin{equation}\label{eq: averaged_phase_diff_equation_}
\dot{\vartheta_1} - \dot{\vartheta_2} = % - \frac{\epsilon}{2} \left(\frac{A_2}{A_1} + \frac{A_1}{A_2}\right) \sin(\vartheta_1 - \vartheta_2 - \Theta_{12}) \nonumber \\
  \mathfrak{A}(t) \sin (\vartheta_1 - \vartheta_2 - \Theta_{12}),
\end{equation}
where, $\mathfrak{A}(t)= {\epsilon} \left({A_2}/{A_1} + {A_1}/{A_2}\right)/2$. The above argument is adapted from \cite{PhysRevLett.76.1804}.  As the amplitudes $A_{1,2}$ are assumed to be slower and in the time scales of $\vartheta_{1,2}$ can be assumed to be constants, from which it follows that Eq.(\ref{eq: averaged_phase_diff_equation_}) has a stable fixed point $\Theta_{12}$.
\begin{figure}[htbp]
    \centering
    % First subfigure
    \begin{subfigure}{0.45\textwidth}
        \centering
\includegraphics[width=0.99\linewidth]{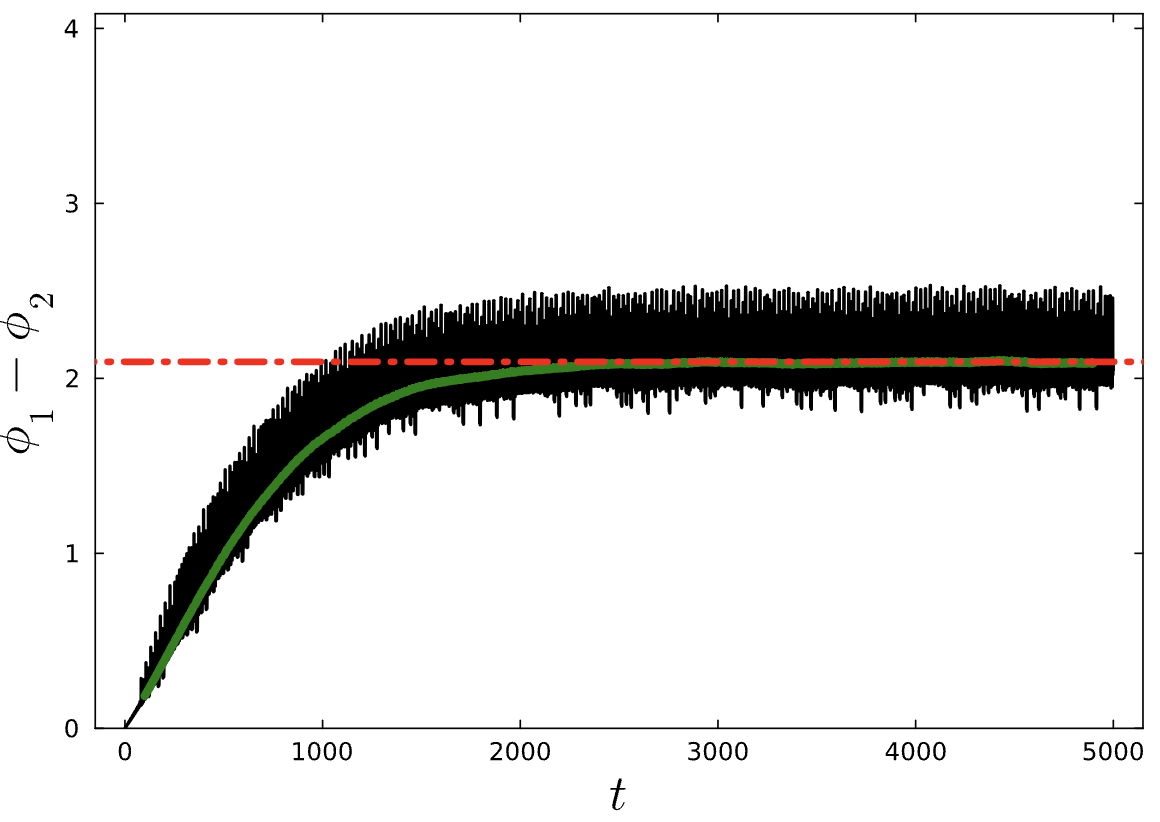}%  
        \caption{}
    \end{subfigure}
    \hfill
    % Second subfigure
    \begin{subfigure}{0.45\textwidth}
        \centering
\includegraphics[width=0.99\linewidth]{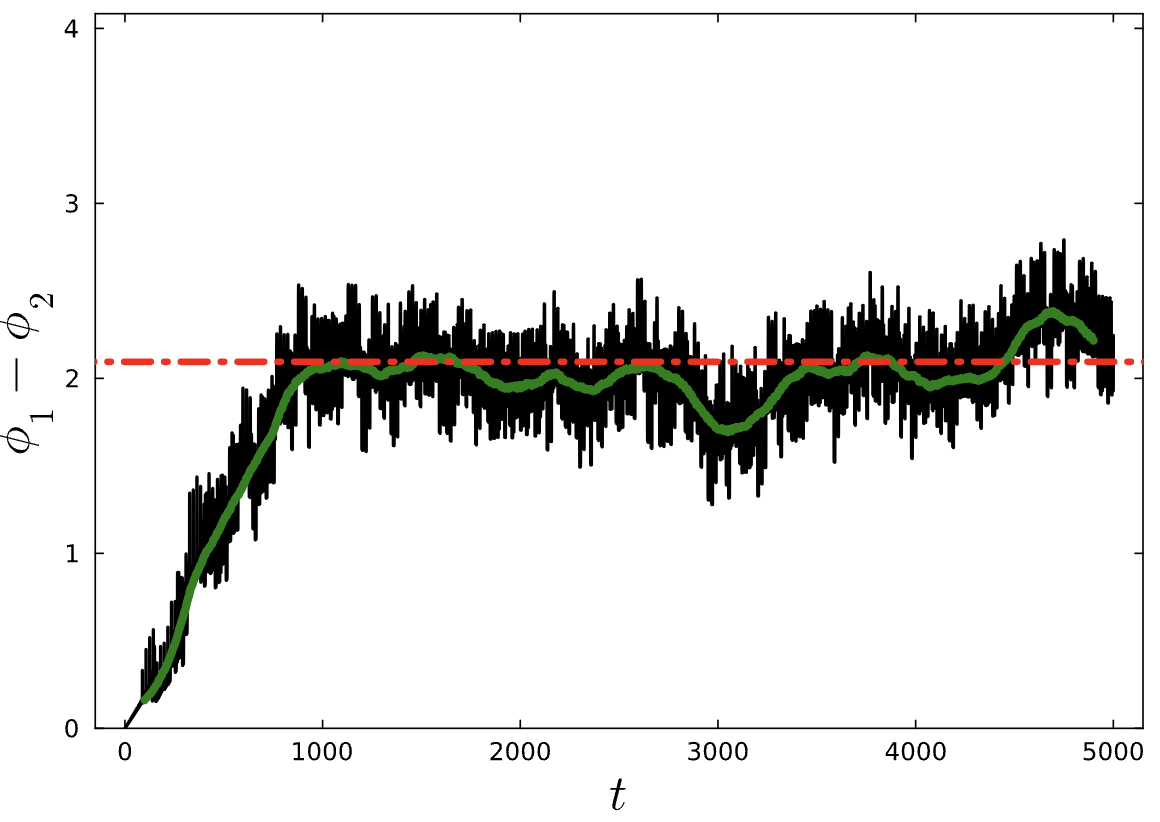}
        \caption{}
    \end{subfigure}
    \caption{Time evolution of phase difference for two coupled R\"ossler systems at (a) sparsely chaotic regime ($\mathfrak{c}=9$) and (b) strongly chaotic regime($\mathfrak{c}=14$). Black full line: numerical simulations; Green full line: Temporal moving average; red dash-dot line: Target phase difference $\Theta = 2\pi/3$. Identical initial values were used in both cases. Numerical values of parameters for Eq.(\ref{eq:CAM_with_second_order_coupling}):  $N=2$, $K=1$, $\mathfrak{a}=0.1$, $\mathfrak{b}=0.1$, $r = 0.001$.}  
    \label{fig : plot_phase_difference}
\end{figure}
This result is  numerically validated through Fig.\ref{fig : plot_phase_difference}, which shows the numerical simulations for two different chaotic regimes. The horizontal dash-dot line at $\Theta_{12} = 2\pi/3$ in Figs. \ref{fig : plot_phase_difference} (a,b) is the desired phase difference. The time varying fluctuations is obtained from direct numerical simulations. Fig.\ref{fig : plot_phase_difference}(a) corresponds to sparse chaotic regime when the R\"ossler parameter $\mathfrak{c} = 9$. While the time history obtained from numerical simulations show fast fluctuations, the temporal moving average as shown the by the full line, is seen to converge close to the desired value. 
%The time history of the phase difference is seen to converge close to the desired value - as seen by the temporal moving average - but continues to oscillate around it. 
Fig.\ref{fig : plot_phase_difference}(b) shows the results for a highly chaotic regime, when  $\mathfrak{c}=14$.
%shows the corresponding results when the parameter  $\mathfrak{c}$ is changed to $14$  which corresponds to a highly chaotic regime. 
Here too the time history of the phase difference  approaches $\Theta_{12}$, but the moving time averaged mean fluctuates about this value. 

These results offer three key takeaways. First, they demonstrate the ability of coupled chaotic systems to achieve a desired phase difference, which can serve as a fixed point of the system, as shown in Eq.(\ref{eq: averaged_phase_diff_equation_}).  Second, the findings support the proposed definition of phase in chaotic systems, lending credence to the possibility of memory encoding in COAM via phase-synchronized states. %These numerical simulations demonstrate two things. 
%First is the ability of coupled chaotic systems to reach a desired state of phase difference; this can be the fixed point of the system as shown through Eq.(\ref{eq: averaged_phase_diff_equation_}). Secondly, these results validate the definition of phase of chaotic systems and provides credence that memory encoding is possible in COAM through phase synchronization states. 
Finally, the absence of amplitude correlation %in the amplitudes of these fluctuations 
and the clear dependence of the magnitude of the phase fluctuations on the underlying chaotic regime, 
%the extent of fluctuations about the desired phase difference on the chaotic regime,
can be attributed to the influence of $A_i$ on the phase $\phi_i$, thereby validating the general form of Eq.(\ref{eq:phase_of_a_general_chaotic_oscillator}).
%As the amplitudes of these fluctuations are uncorrelated, it follows that these can be attributed to the effects of the amplitude $A_i$ on the phase $\phi_i$, and are therefore dependent on the chaotic regime. 

\begin{figure}[htbp]
    \centering
    \includegraphics[width=0.9\linewidth]{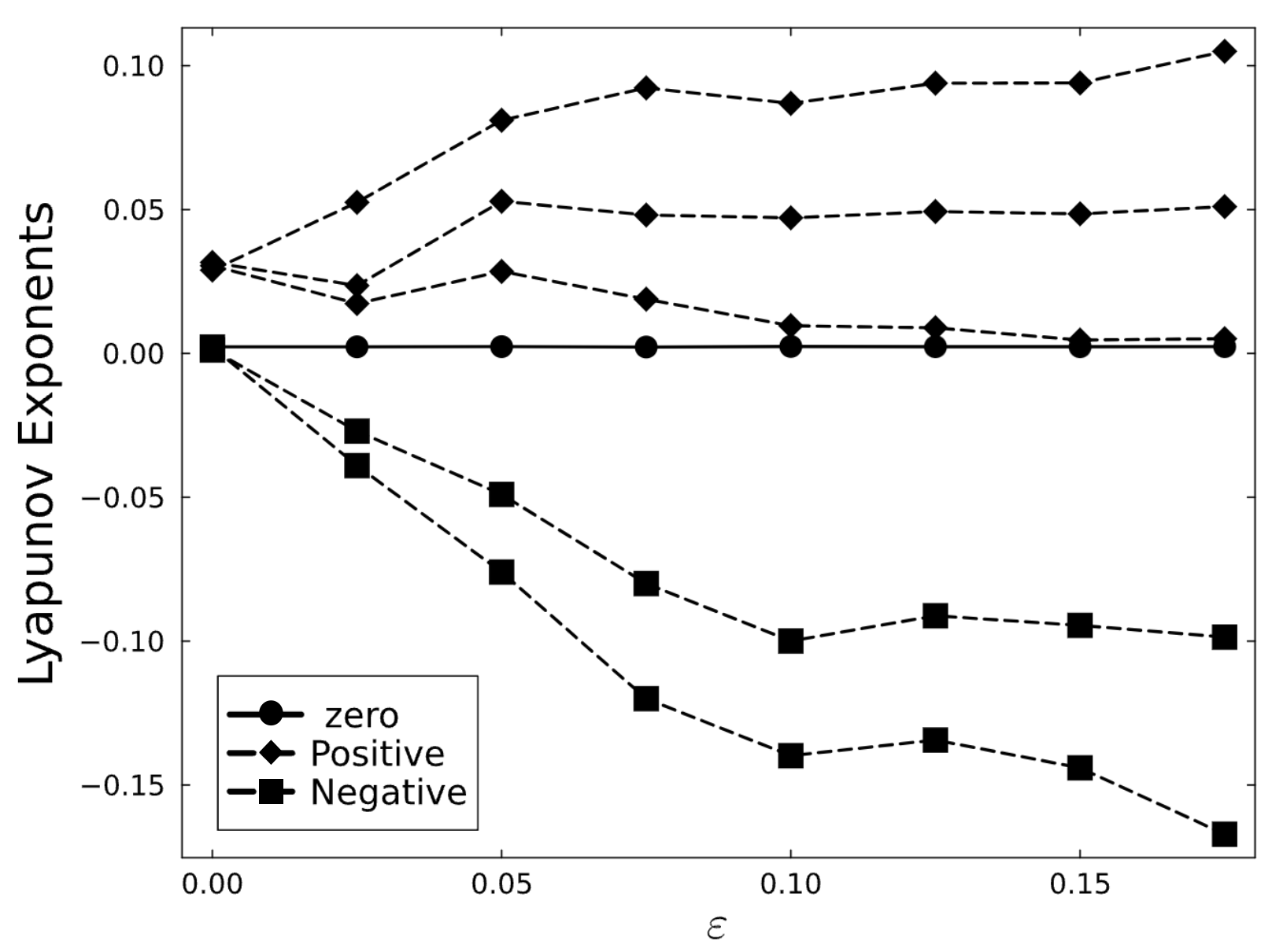}
    \caption{Lyapunov spectrum of a network of 3 coupled R\"ossler systems, as a function of coupling strength $\epsilon$. Three zero LEs are observed for $\epsilon=0$; only one zero LE is seen for $\epsilon >0$. The 6 largest LEs are shown; the remaining 3 LEs are significantly negative to be shown in this scale. The numerical parameters usef in Eq.(\ref{eq:CAM_with_second_order_coupling}) are  $N=3$, $K=1$, $\mathfrak{a} = 0.1$, $\mathfrak{f} = 0.1$, $\mathfrak{c} = 9$.}    
    %The six largest Lyapunov exponents  as functions of the coupling strength $\epsilon$ for Eq. (\ref{eq:CAM_with_second_order_coupling}) with $N=3$ and $K=1$. Phase synchronization is instantaneously established, as indicated by the immediate shift of two zero Lyapunov exponents to negative values upon introducing coupling. The remaining three negative LEs are orders of magnitude high negative, and are ignored in the plot. The parameters are set to $\mathfrak{a} = 0.1$, $\mathfrak{f} = 0.1$, $\mathfrak{c} = 9$, and the weight matrix is defined according to Eq. (\ref{eq:phase_weight}).}
    \label{fig:N_4_largest_lyapunov_exponent}
\end{figure}

Confirming the existence of phase synchronization in networks with larger number of coupled chaotic oscillators is carried out numerically by computing the Lyapunov spectrum of the system. As is well known, Lyapunov exponents (LEs) quantify the long-term dynamical behavior of system trajectories. Uncoupled R\"ossler systems in the chaotic regime exhibit one positive LE, one negative LE and one zero LE; the zero LE corresponds to translation along the trajectory and represents phase shift \cite{PhysRevLett.76.1804}. %The next question addressed is the existence of phase synchronization in networks with a larger number of coupled chaotic systems. This can be numerically investigated by computing the Lyapunov spectrum of the system. 
For a network of $N$ coupled R\"ossler systems, there are $3N$ LE. In a phase-synchronized state, the phases of all oscillators become linear translations of one another implying that only one zero LE should remain in the spectrum.   The coupling effectively converts $N-1$ zero LEs into negative LEs, resulting in a total of $2N-1$ negative LEs. This  reflects an attractive interaction between the phases, effectively enforcing a constant phase difference between the oscillators. As a result, a synchronized system is expected to exhibit $2N-1$ negative LEs, one zero LE, and $N$ positive LEs. Numerical validation of this behavior is shown for the case $p=1$, $K=1$ and $N=3$ in Eq.(\ref{eq:CAM_with_second_order_coupling}).  
 %The next question that is addressed is establishing the existence of phase synchronization in a network with larger number of coupled chaotic systems. This can be numerically established by computing the Lyapunov spectrum for the system. As is well known, Lyapunov exponents (LE)s are quantifiers that enable characterizing the long term dynamical features of the system states. Individual R\"ossler systems in the chaotic regime have one positive LE, one negative, and the third is zero. The zero LE corresponds to translation along the trajectory and represents the shifting of the phase \cite{PhysRevLett.76.1804}. For a network of $N-$coupled R\"ossler systems, there exists $3N$ LE. However, the state of phase synchronization implies that the phases in all the oscillators in the network are linear translations of any of them and therefore the system is expected to have only one zero Lyapunov exponent (LE). The coupling therefore converts $N-1$ zero LEs into negative LEs, resulting in a total of $2N-1$ negative LEs. This transformation reflects an attractive interaction between the phases, effectively enforcing a constant phase difference between the oscillators. The remainder of the spectrum consists of $N$ positive LEs and one zero LE. 
  %The presence of only one zero LE in a CAM  is numerically validated for the case when $p=1$, $K=1$ and $N=3$. 
Fig.\ref{fig:N_4_largest_lyapunov_exponent} shows the Lyapunov spectrum for the network as a function of the coupling strength $\epsilon$.  For $\epsilon=0$ (the uncoupled case), the spectrum includes three zero LEs, as expected. However, for $\epsilon>0$ 
only one zero LE is observed, indicating that even weak coupling is sufficient to induce phase synchronization. These observations have been found to hold for networks with larger values of $N$.

A theoretically more rigorous approach to demonstrate that the network achieves phase synchronization is by showing the existence of an energy function.  From the perspective of dynamical systems theory, a stable state corresponds to a minima of an underlying energy landscape. As previously discussed, the phase-synchronized state in a COAM exhibits  fluctuations around a mean phase difference,  which deviates from traditional interpretations of stable states defined strictly by minima in energy functions. %is not completely in sync with the traditional energy based definitions of a stable state. 
Nevertheless, an energy function associated with the system in Eq.(\ref{eq:CAM_with_second_order_coupling})  can be formulated as %expressed as $\mathcal{L}(\mathbf{\phi; W}, \epsilon, \epsilon_k) =$ 
\begin{eqnarray}
\mathcal{L}(\mathbf{\phi; W}, \epsilon, \epsilon_k) & = &     \mathbf{-} \Re\left(\sum_{i,j = 1}^N W_{i,j} e^{\mathrm{i}\phi_i}e^{-\mathrm{i}\phi_j}\right) \nonumber \\
& &- \sum_{k=2}^K\Re\left(\frac{\epsilon_k}{4N}\sum_{i,j = 1}^N e^{\mathrm{i}~k\phi_i}e^{-\mathrm{i}~k\phi_j}\right).
    \label{eq:OAM_2nd_Lyapunov_CAM}
\end{eqnarray}
The existence of this energy function ensures that in an averaged sense, any solution converges to a phase-locked state as $t\rightarrow\infty$. 
\begin{figure}[htbp]
    \centering
    \includegraphics[width=0.9\linewidth]{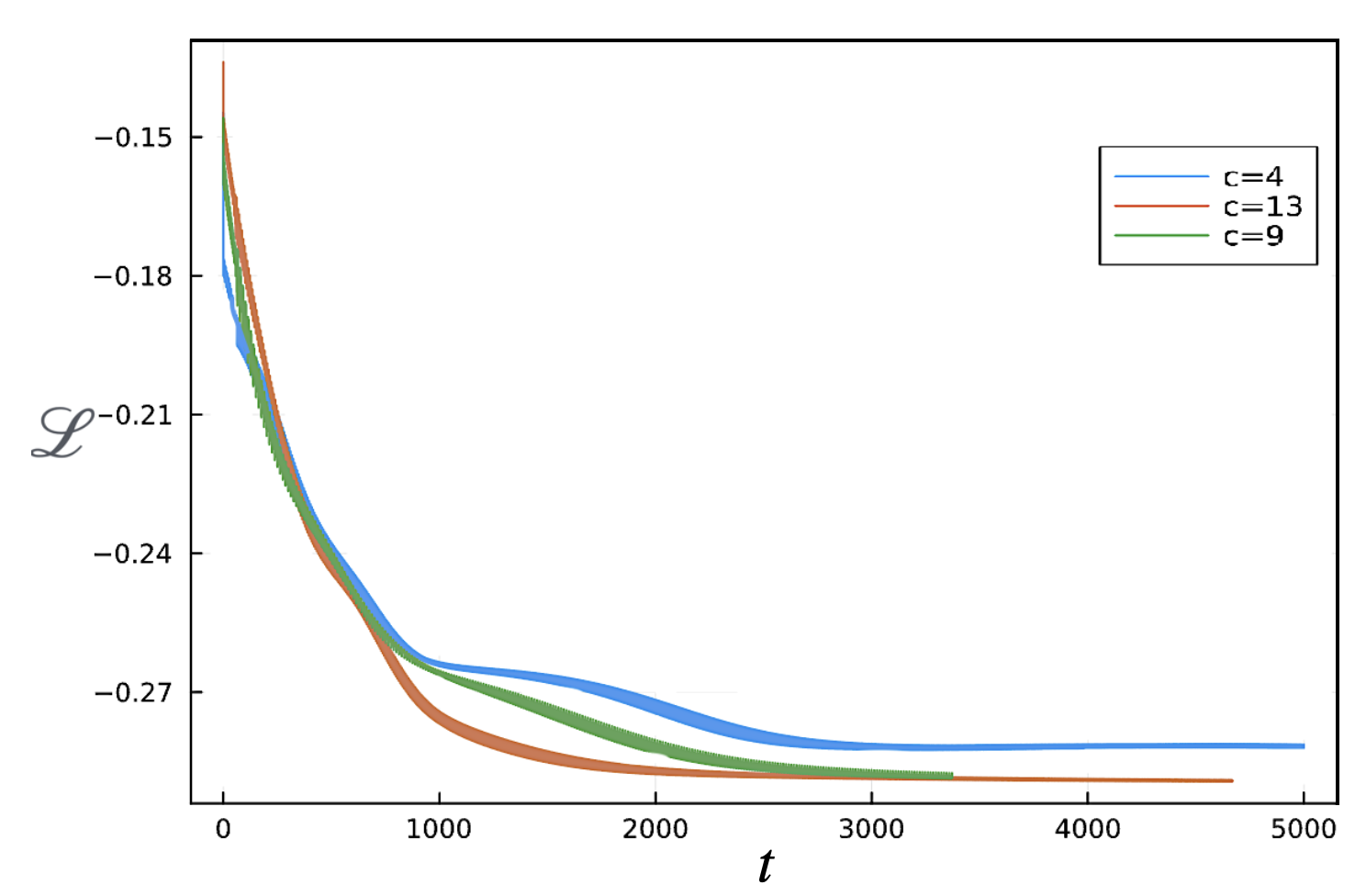}
    \caption{Time evolution of the energy function for the COAM; the   underlying R\"ossler systems correspond to periodic ($\mathfrak{c}=4$), sparsely chaotic ($\mathfrak{c}=9$) and strongly chaotic ($\mathfrak{c}=14$) regimes. The numerical values of the parameters for Eq.(\ref{eq:CAM_with_second_order_coupling}) are  $N=50$, $\mathfrak{a} = \mathfrak{f} = 0.1$, $\epsilon = 0.001$.}
    %in time for three different $\mathfrak{c}$ values. The R\"{o}ssler system is in periodic regime for $\mathfrak{c } = 4$ and in sparse chaotic regime for $\mathfrak{c}=9, 13$.  The Lyapunov function evolution shows fast instantaneous oscillations, but the average of the Lyapunov function approaches steady state with time. The weight matrix is initialized using 1 binary patterns for $N=50$. Other parameters were set at $\mathfrak{a} = \mathfrak{f} = 0.1$, $\epsilon = 0.001$}
    \label{fig:lyapunov_function}
\end{figure}
Fig. \ref{fig:lyapunov_function} illustrates a numerical simulation of the evolution of the energy function over time for the case $K=1$, $N=50$, $\mathfrak{a} = 0.1$, $\mathfrak{b} = 0.1$,  $\epsilon = 0.001$, and $\mathfrak{c} = 4, 9, 14$. The weight matrix $\bf W$ is initialized for $p=1$ as per Eq.(\ref{eq:phase_weight}). For all $\mathfrak{c}$ values, the energy function exhibits a decreasing trend over time, while still oscillating near a fixed point, thereby confirming the claim of the network achieving a stable phase-synchronized state.

\subsection{Memory retrieval with only first Fourier mode coupling }
\begin{figure}[htbp]
    \centering
    \includegraphics[width=0.9\linewidth]{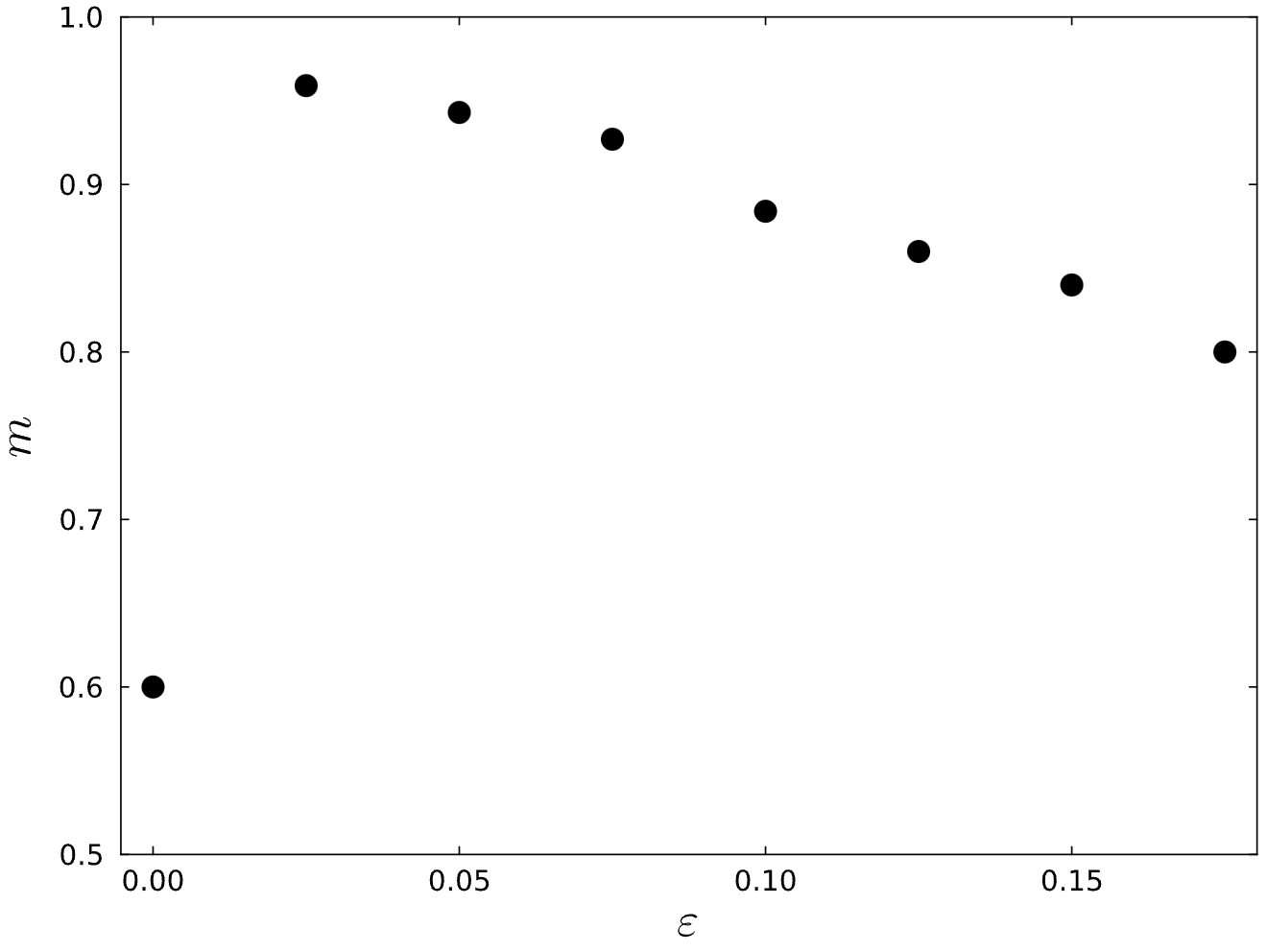}
    \caption{The quality of memory retrieval  in terms of average overlap measure $m^\eta$ as a function of coupling strength $\epsilon$. The numerical values of the parameters for Eq.(\ref{eq:CAM_with_second_order_coupling})  are $N=100$, $K=1$,$\mathfrak{a} = 0.1$, $\mathfrak{f} = 0.1$, $\mathfrak{c} = 9$.}
    
    %The overlap $m$ is maximized for $\epsilon < 0.05$; beyond this range, $m$ steadily decreases. The parameters are set to $\mathfrak{a} = 0.1$, $\mathfrak{f} = 0.1$, $\mathfrak{c} = 9$, and the weight matrix is defined according to Eq. (\ref{eq:phase_weight}).}
    \label{fig:overlap_p_1}
\end{figure}

While the results in the previous section confirm the existence of a phase synchronization state in COAM,  the question of whether synchronization  occurs at the desired state remains to be addressed. The desired state is encoded via the weight matrix ${\bf W}$ and the quality of memory retrieval is quantified using the order parameter $m$; see Eq.(\ref{eq:continuous_overlap_m}). Fig.\ref{fig:overlap_p_1} illustrates the variation of $m$ as a function of the coupling strength $\epsilon$ for the COAM model with $K=1$, with only a single stored pattern. 
The system is initialized with a partial memory,  having an overlap of $60\%$ with the stored pattern; this is the value of $m$ for $\epsilon=0$. As shown in Fig.\ref{fig:overlap_p_1}, the value of  $m$ rises to approximately $0.96$ for small $\epsilon$, indicating that the network not only achieves phase  synchronization but does so at the desired state, resulting in high-quality memory retrieval. However, as the coupling strength increases, a decline in $m$ is observed reflecting a deterioration in retrieval quality. This suggests that while the system maintains phase synchronization, it transitions to a state with only partial overlap with the desired memory.   Similar trends are observed across different dynamical regimes, characterized by variations in the parameter $\mathfrak{c}$. These observations underscore the importance of operating COAM models in the weak coupling regime, where the trade-off between synchronization and memory fidelity is optimized.

%to deteriorate with increasing coupling strength, as denoted by the downward trend in the figure. The dynamical implication of this is that with stronger coupling, the system achieves phase synchronization but at a state that has partial overlap with the desired state. Similar trends are observed for all regimes, characterized by the parameter $\mathfrak{c}$. This motivates considering only weak coupling in CAM models. 

In the case of multiple stored patterns, the retrieval dynamics of COAM exhibit interesting features that strongly depend on the chaotic regime of the oscillators. %Fig.\ref{fig:transient_time_evolution_order_parameter} compares 
The time evolution of the order parameter $m$ are investigated for two distinct dynamical regimes: periodic ($\mathfrak{c}=4$) and chaotic ($\mathfrak{c}=14$). 
\begin{figure}[htbp]
    \centering
     \includegraphics[width=0.9\linewidth]{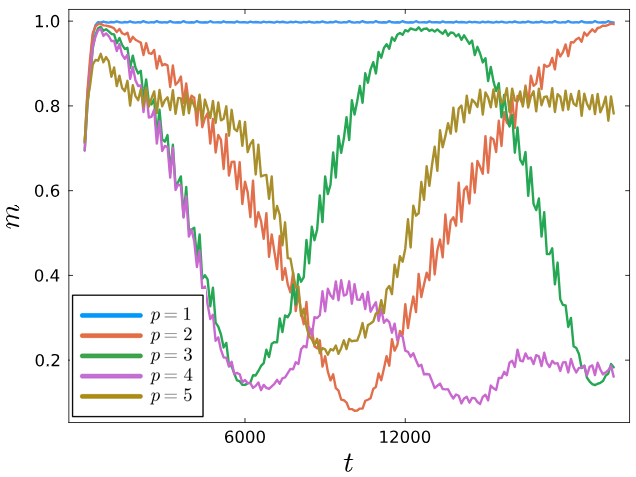}
 %   \hfill
    \caption{Retrieval quality of memories, quantified through overlap measure $m$ as a function of time for COAM with the  R\"ossler systems in the periodic ($\mathfrak{c}=4$) regime,for cases corresponding to number of stored memories varying from $p=1$ to $p=5$. Initial overlap with one of the stored patterns is 0.7. Numerical values for the parameters for Eq.(\ref{eq:CAM_with_second_order_coupling}): $N = 100$, $K = 1$, ($\mathfrak{a, b} = 0.1$), $\epsilon = 0.001$.} 
    
    %The long-term evolution of the order parameter $m$ is investigated for increasing $p$ at two different values of $\mathfrak{c}$, $\mathfrak{c}=4$ with a fixed coupling strength $\epsilon = 0.01$. The weight matrix is constructed following Eq. (\ref{eq:phase_weight}). The model is initialized with a partial overlap with any of the stored patterns $m^{(\eta)}_{0} \approx 0.7$, for all $p$ values, and the equations are solved over time with $m$ recorded at each time step. For $\mathfrak{c}=4$, wherein each R\"ossler system is in periodic regime, for $p=1$, $m$ increases and converges to unity, demonstrating perfect retrieval; however, with $p=2, 3$, the network oscillates between two stored patterns, while for $p=4, 5$ the network initially reaches high overlap ($m \approx 1$) before decreasing, indicating instability of the stored state. }
\label{fig:transient_time_evolution_order_parameter}
\end{figure}
For both regimes, the time evolution of $m$ is shown when up to 5 patterns are stored in the network, and the system is initialized with a partial memory that has approximately $70\%$ overlap with one of the stored patterns. At each time step, the overlap is computed with this target pattern, allowing an assessment of how the system evolves with respect to the desired memory. 
%In all cases, the model is initialized with a partial overlap with one of the stored patterns, such that the initial overlap is approximately $70\%$. The time histories of $m$ are plotted to investigate the convergence characteristics towards the desired memory states, and is carried out by computing the overlap with the stored pattern at each time step. 
Fig.\ref{fig:transient_time_evolution_order_parameter}, which corresponds to periodic regime, shows very interesting features. For a single stored pattern ($p=1$), $m$ increases monotonically and attains a steady state value close to unity indicating near-perfect memory retrieval. When the number of stored patterns is increased to $p=2$, the system initially exhibits similar behavior, with $m$ rising rapidly close to unity. However, as the simulation is extended in time, $m$ begins to decrease and enters a periodic regime where it oscillates between the target pattern and the second stored pattern. At the minima of this oscillation, the overlap with the second pattern approaches unity, demonstrating that the system alternates between the two memory states over time. This switching behavior persists even for longer durations, and a similar phenomenon is observed for $p=3$, where the system oscillates between the desired pattern and one of the two other stored patterns. Interestingly, no specific correlation could be established to predict which of the other two patterns the system tends to oscillate toward. As the number of stored patterns increases to $p=4$, the overlap $m$ reaches a maximum value that is lower than in the earlier cases, indicating poorer retrieval. Furthermore, the system progressively diverges from the desired memory, and while some periodicity remains in the evolution of $m$, the amplitude of subsequent maxima diminishes significantly, suggesting an inability to return to the target memory. The retrieval quality continues to deteriorate with additional patterns, leading to the inference that the desired state becomes unstable in this dynamical regime when more than one pattern is stored. This observation is consistent with prior results from OAM models, which have shown that in periodic regimes, stable phase-synchronized states are achievable only when the number of stored patterns does not exceed two.

\begin{figure}[htbp]
    \centering
    %\centering
     %\includegraphics[width=0.9\linewidth]    %\begin{subfigure}[b]{0.45\textwidth}
      %  \centering
        \includegraphics[width=0.95\linewidth]{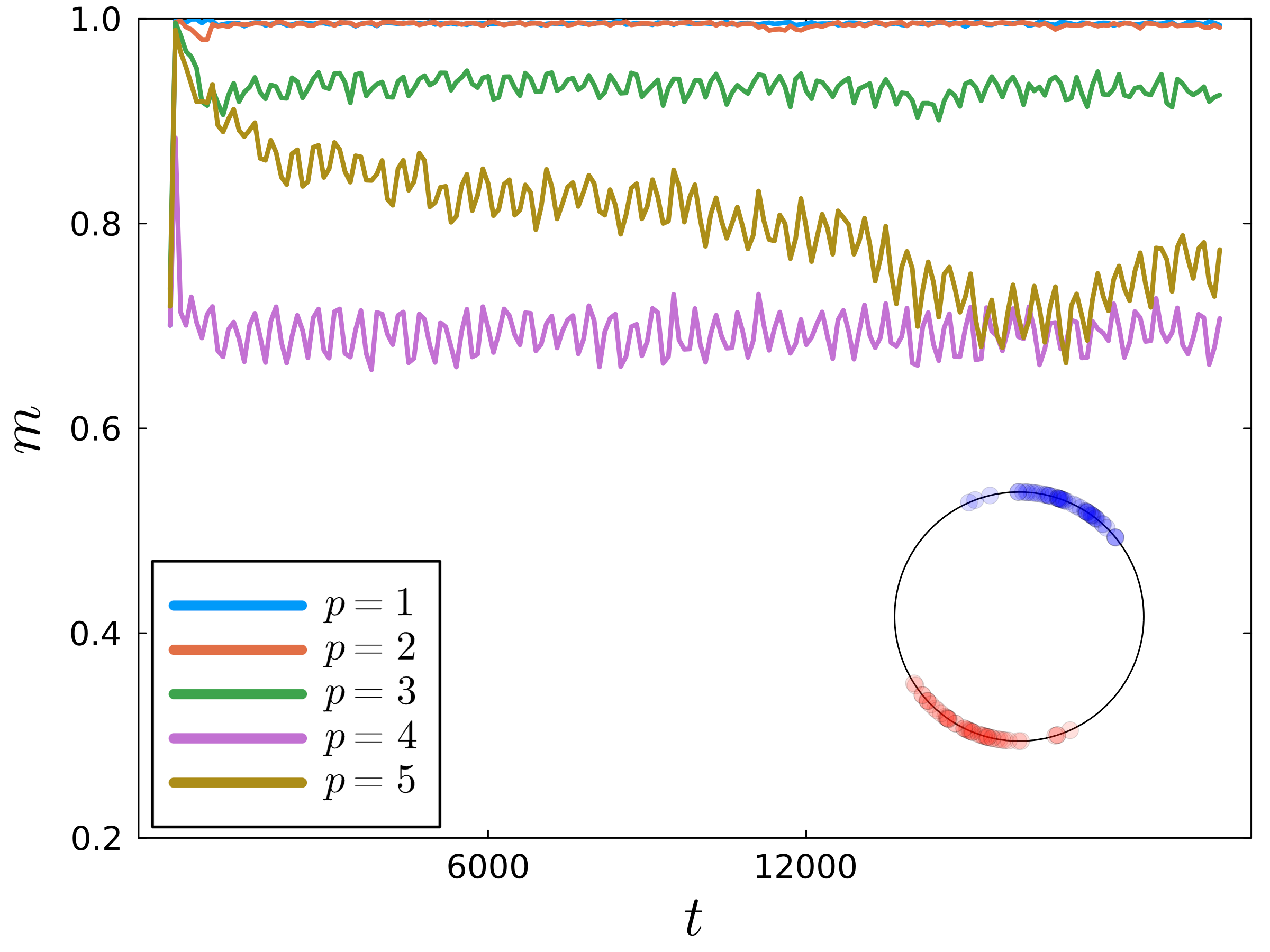}%
    \caption{Retrieval quality of memories, quantified through overlap measure $m^\eta$ as a function of time for COAM with the R\"ossler systems in the strongly chaotic ($\mathfrak{c} = 14$) regime, for cases corresponding to number of stored memories varying from $p=1$ to $p=5$. Initial overlap with one of the stored patterns is 0.7. Numerical values for the parameters for Eq.(\ref{eq:CAM_with_second_order_coupling}): $N = 100$, $K = 1$, ($\mathfrak{a, b} = 0.1$), $\epsilon = 0.001$.  Inset shows the antiphase synchronization clusters of the neuronal phases. }
    \label{fig:overlap_with_alpha_for_c_4_and_18}
    \end{figure}

The retrieval dynamics in the chaotic regime ($\mathfrak{c}=14$) are markedly different, as illustrated in Fig. \ref{fig:overlap_with_alpha_for_c_4_and_18}. 
    For all values of $p$, the order parameter $m$ initially reaches values close to unity, indicating the system's ability to rapidly achieve synchronization. However, unlike in the periodic case, no oscillations between stored memory states are observed for larger $p$. Instead, the system converges to a steady state. While this indicates the presence of stable phase-synchronized dynamics, the final states do not correspond to the desired memory, indicating  poor retrieval performance. Thus, while chaos appears to suppress instabilities and stabilize phase synchronization, it does so at the cost of accurate memory recall.

\begin{figure}[htbp]
    \centering
   % \begin{subfigure}[b]{0.45\textwidth}
   %     \centering
   %     \includegraphics[width=0.95\linewidth]{Figures/transient_plots_p_1_5_20000_c_14.png}%
   %     \caption{}
  %  \end{subfigure}
  %  \hfill
    %\begin{subfigure}[b]{0.45\textwidth}
       % \centering
          \includegraphics[width=0.93\linewidth]{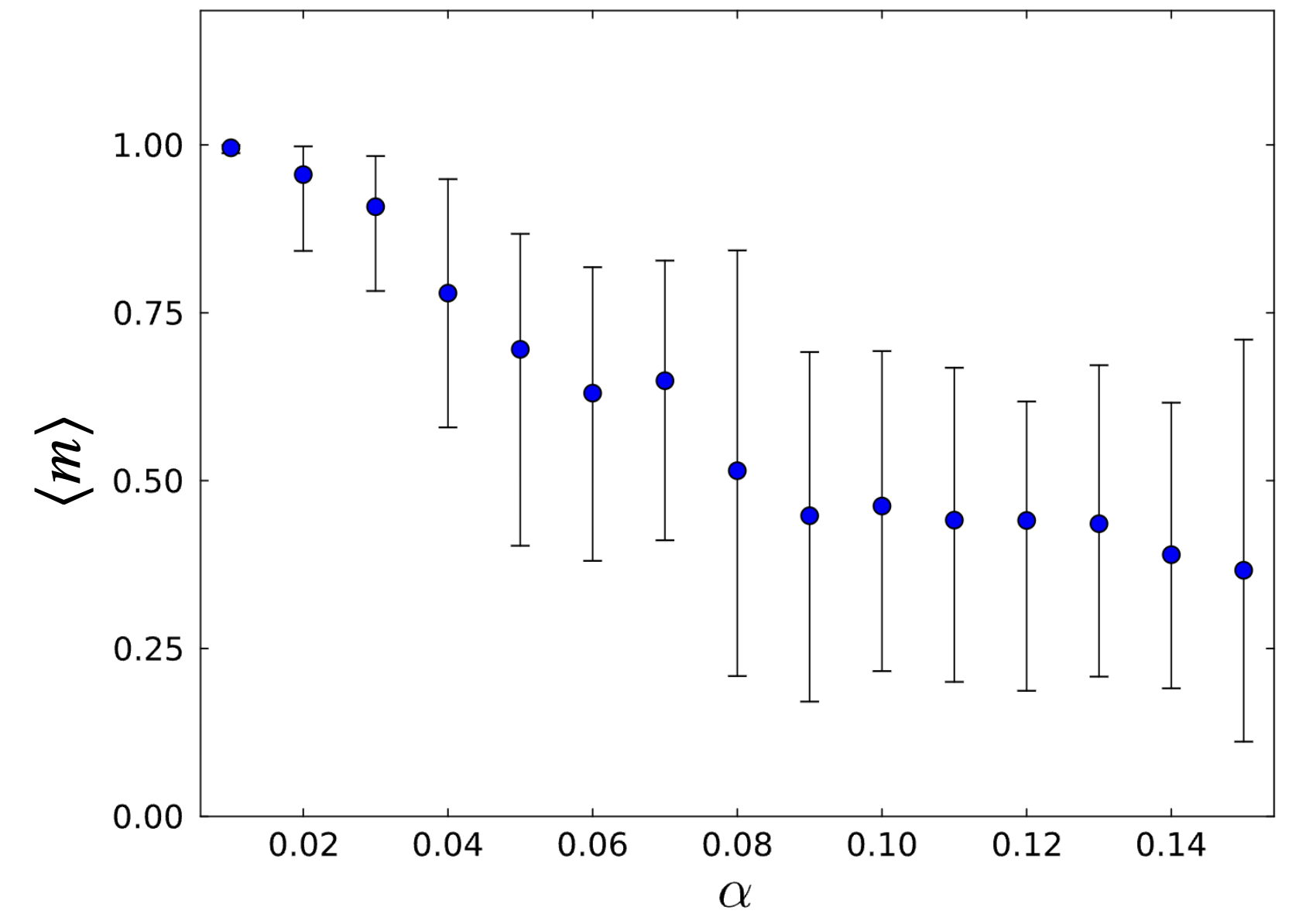}%       
     %   \caption{}
    %\end{subfigure}
    \caption{Average overlap $<m>$ as a function of loading parameter $\alpha$; circles: mean values, vertical lines represent the variation between the maximum and minimum values.   Numerical values for the parameters for Eq.(\ref{eq:CAM_with_second_order_coupling}) are: $N=100$, $\mathfrak{a}=\mathfrak{b}=0.1$.}   
   % The long-term evolution of the order parameter $m$ is investigated for increasing $p$ at two different values of $\mathfrak{c}$, $\mathfrak{c}=14$ with a fixed coupling strength $\epsilon = 0.01$. The weight matrix is constructed following Eq. (\ref{eq:phase_weight}). The model is initialized with a partial overlap with any of the stored patterns $m^{(\eta)}_{0} \approx 0.7$, for all $p$ values, and the equations are solved over time with $m$ recorded at each time step. For all the cases, $m$ attains a maxima close to unity  but for cases $p>2$ at longer time evolutions, the system converges to a steady state without exhibiting the oscillations between stored patterns as seen in the previous case. However, the steady phase synchronized states are different from the desired states indicating that while the phase synchronized states are stable, the quality of memory retrieval is poor. The distribution of phases have been for the case $p=2$ is shown in the inset. The two colors correspond to the two distinct clusters formed as a result of antiphase synchronization.
   % (b) The average overlap $ m $ is shown as a function of $ \alpha $ for $ \mathfrak{c} = 4 $ (left) and $ \mathfrak{c} = 14 $ (right). The average overlap is computed as per Algorithm-1. The vertical lines represent error bars, indicating the maximum and minimum overlap values. The parameters used are $ N = 100 $, $ \mathfrak{a} = 0.1 $, and $ \mathfrak{b} = 0.1 $.}
    \label{fig:overlap_with_alpha_for_c_4_and_18b}
\end{figure}

To examine these retrieval dynamics more generally, further simulations were conducted using multiple sets of randomly generated memory patterns. These simulations also accounted for variations in the initial conditions, specifically the initial values of the amplitude $A_i$  and the third state  variable $z_i$, as detailed in SI Appendix B. The quality of retrieval was quantified through the average overlap $\langle m\rangle$ and its statistical spread. Fig. \ref{fig:overlap_with_alpha_for_c_4_and_18b} presents the results as a function of the loading parameter $\alpha$, which denotes the ratio of stored patterns to network size. 
It is observed that reliable retrieval occurs only for $\alpha \leq 0.04$, consistent with the storage capacity of $0.04$ previously reported for OAMs with single Fourier mode coupling. These results suggest that while chaotic dynamics may confer robustness to phase synchronization, they inherently limit the system’s retrieval accuracy.

\begin{figure}[htbp]
    \centering
    \includegraphics[width=0.9\linewidth]{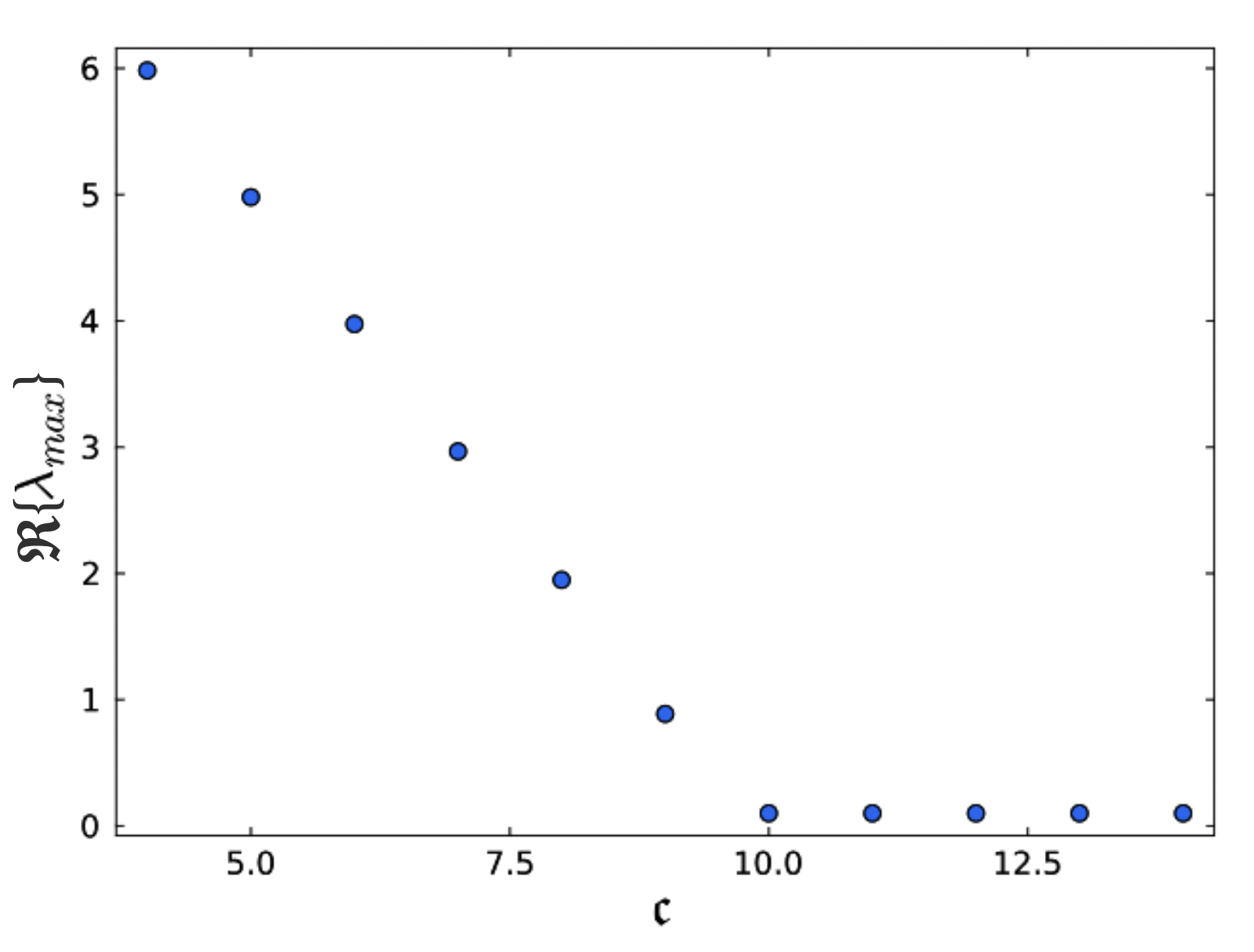}
    \caption{The maximum eigenvalue $\text{Re}~\lambda_{\max}$ of the Jacobian, linearized at $\mathfrak{b}$ for $\alpha = 0.06$, is plotted as a function of $\mathfrak{c}$. The other state variables are initialized using the system's ensemble averages. The coupling strength is fixed at $\epsilon = 0.001$, with the other numerical values for the parameters for Eq.(\ref{eq:CAM_with_second_order_coupling}) being $N = 100$, $\mathfrak{a} = 0.1$, and $\mathfrak{b} = 0.1$. As $\mathfrak{c}$ increases, $\text{Re}~\lambda_{\max}$ decreases.}
    \label{fig:lambda_max_second_coupling}
\end{figure}

This can indeed be validated by examining the stability of the stored patterns across different dynamical regimes.  In traditional OAMs, stability can be conveniently assessed using the largest eigenvalue of the Jacobian matrix derived from the linearized system around a stored pattern. In COAM, however, this analysis becomes more complex since the Jacobian depends not only on the phases but also on the instantaneous values of $A_i$ and $z_i$.  To facilitate this, the Jacobian corresponding to Eq.(\ref{eq:CAM_with_second_order_coupling}) is evaluated at one of the stored patterns, with $A_i$ and $z_i$ initialized using their ensemble averages. Due to the asymmetry of the Jacobian, its eigenvalues are generally complex, and the stability of the stored pattern is determined by the largest real part of these eigenvalues, denoted by $\Re\{\lambda_{\max}\}$.
%For OAMs, this can be easily investigated in terms of the largest eigenvalue of the Jacobian obtained by linearizing the system at the stored patterns. However, in CAM, the eigenvalues of the Jacobian of the corresponding linearized system depend not only on the phases but also on the state variables $A_i$ and $z_i$. The Jacobian corresponding to Eq.(\ref{eq:CAM_with_second_order_coupling}) is evaluated at one of the stored patterns, with the state variables $A_i$ and $z_i$ being initialized with their ensemble averages. %The other parameters are fixed at $\epsilon = 0.001$ and $\epsilon_2 = 0$.
%As the Jacobian is asymmetric, its eigenvalues are complex and the largest real part of the eigenvalues, denoted by $\Re\{\lambda_{\max}\}$ determines the stability of the stored patterns. 
Fig.\ref{fig:lambda_max_second_coupling} shows the variation of $\Re\{\lambda_{\max}\}$ as  a function of $\mathfrak{c}$. As $\mathfrak{c}$ increases, indicating a transition to more chaotic behavior in the uncoupled R\"ossler oscillators, a decreasing trend in $\Re\{\lambda_{\max}\}$ is observed. This indicates that stored patterns become progressively less unstable as the level of chaos increases, thus potentially improving the system’s retrieval capacity. However, it is noteworthy that $\Re\{\lambda_{\max}\}$ remains positive throughout, reaffirming the residual instability of these patterns.  Additionally, the form of the coupling function plays a crucial role in determining retrieval quality. This function depends instantaneously on the amplitude $A_i$ and evolves over time in either a periodic or chaotic manner, depending on the operating regime of the individual oscillators. At higher levels of chaos, the phase difference between oscillators exhibits larger deviations from the ideal phase-locked states, which further contributes to inaccuracies in retrieval. Interestingly, previous studies \cite{aonishi} have suggested that these deviations—interpreted as temporal ``gaps” in the coupling function - can enhance memory retrieval under certain conditions.

%Note that as $\mathfrak{c}$ increases, the uncoupled  R\"ossler systems moves towards higher chaotic regimes. The trend of decrease in the value of  $\Re\{\lambda_{\max}\}$ with higher values of $\mathfrak{c}$ is indicative that the stored patterns become less unstable with increasing chaos, implying better retrieval. It is worth noting that  $\Re\{\lambda_{\max}\} $ remains positive. Moreover, the coupling function depends instantaneously on the amplitude as described in previously. The coupling function exhibits temporal variability, with its evolution being either periodic or chaotic, depending on whether each individual Rössler system operates in a periodic or chaotic regime. Moreover, the time evolution of phase difference exhibits significant deviations from the desired steady state phase difference for higher chaos. These gaps in the coupling function have been shown to enhance retrieval performance \cite{aonishi}.

\subsection{Memory retrieval with higher Fourier mode coupling}
Memory retrieval performance in COAM is next investigated when higher order Fourier coupling modes are introduced into the model. The COAM model comprising of R\"ossler oscillators (Eq.(\ref{eq:CAM_with_second_order_coupling})) with $N=100$, $K=2$ is considered.  The network is trained by storing three ($p=3$) images from the widely used MNIST dataset \cite{deng2012mnist}.
%Since the phase variable associated with CAM follows similar characteristic features as of OAMs, it is worth examining the memory retrieval when higher order Fourier modes are introduced into the coupling. This is first examined using a CAM model with second order Fourier mode in the phase coupling, and storing three images from the widely used MNIST dataset \cite{deng2012mnist}. 
For the sake of easing computational burden, these images are first reduced from $ 28 \times 28 $ pixels to $ 10 \times 10 $ using bicubic interpolation, followed by thresholding to obtain a binary vector of dimension $100$ and subsequently stored in the model through the  weight matrix in Eq.(\ref{eq:phase_weight}). To test the memory retrieval quality, the network is stimulated with each of these three patterns with partial overlap of $60\%$.
%dimensionality reduction is applied to the image dimensions, reducing them from $ 28 \times 28 $ pixels to $ 10 \times 10 $ using bicubic interpolation, followed by thresholding to obtain a binary vector. Each $ 10 \times 10 $ matrix is converted into a one-dimensional vector of length 100 and are stored in the CAM model through the weight matrix in Eq. (\ref{eq:phase_weight}). As before, the network is initialized with each of these three patterns with partial overlap of $60\%$. 
\begin{figure}[htbp]
    \centering
    \includegraphics[width=0.99\linewidth]{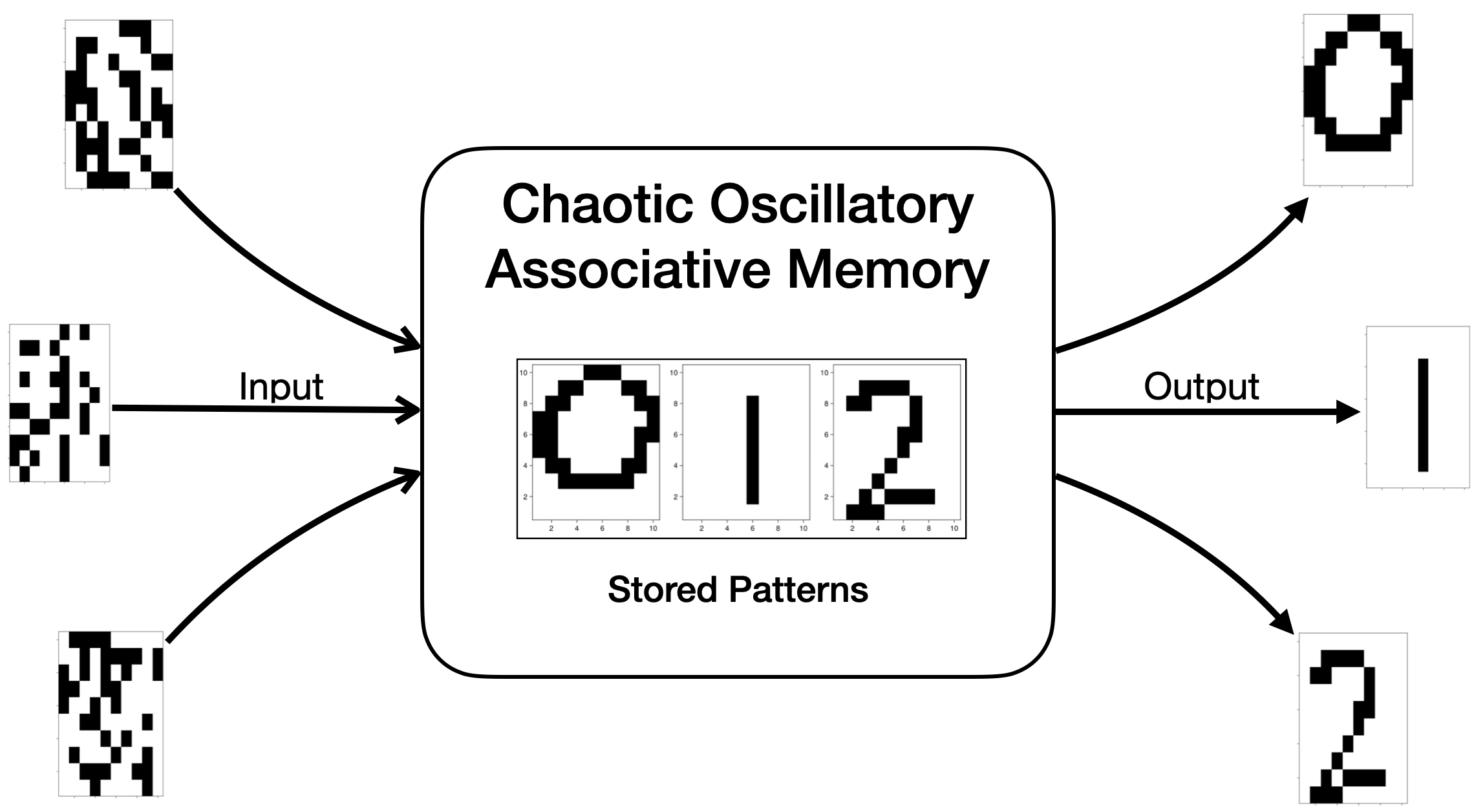}
    \caption{The central block represents the patterns stored in COAM with $N=100$, $K=2$. Images to the left are sample realizations of noisy versions of stored patterns with $m=0.6$, which are used as stimuli. Images on the right are the corresponding retrieved images.} 
    %Left column is a sample realization of each of the stored patterns but embedded with noise such that the overlap is only 60\% i.e., $m = 0.6$. Right column : Retrieved pattern snapshot after binarizing.}
    \label{fig:MNIST_retrieval}
\end{figure}
These input stimuli are shown in Fig. \ref{fig:MNIST_retrieval}  in the left side while the corresponding retrieved images are shown on the right. As can be seen, the retrieval is perfect and is a remarkable improvement from the results presented in the previous section, highlighting the effects of higher Fourier mode couplings in superior memory retrieval. 
%shows a CAM with the three patterns stored. The network initialization with the partial patterns with $60\%$ overlap are shown in the left, while the corresponding retrieved patterns ae shown in the right.  
%A perfect retrieval is observed in all three cases. This is a remarkable improvement from the results reported in the previous section, highlighting that the addition of higher order Fourier modes in the phase coupling enhances the retrieval quality in CAM. It is worth noting that the retrieval quality for $p=3$ is not affected by $\mathfrak{c}$. 

\begin{figure}[htbp]
    \centering
    \begin{subfigure}{0.45\textwidth}
        \centering
        \includegraphics[width=0.99\linewidth]{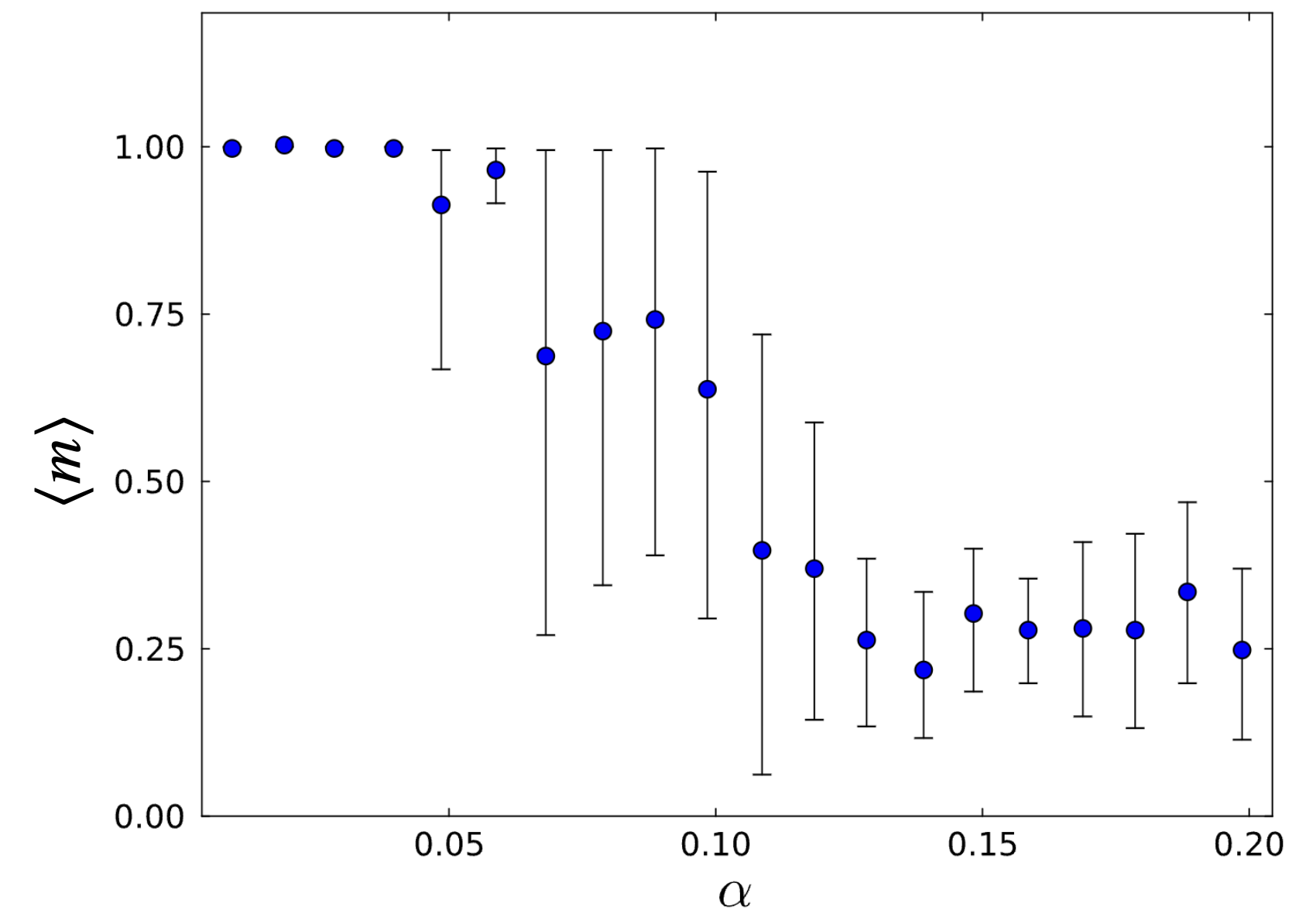}
        \caption{}
    \end{subfigure}
    \begin{subfigure}{0.45\textwidth}
        \centering
        \includegraphics[width = 0.99\textwidth]{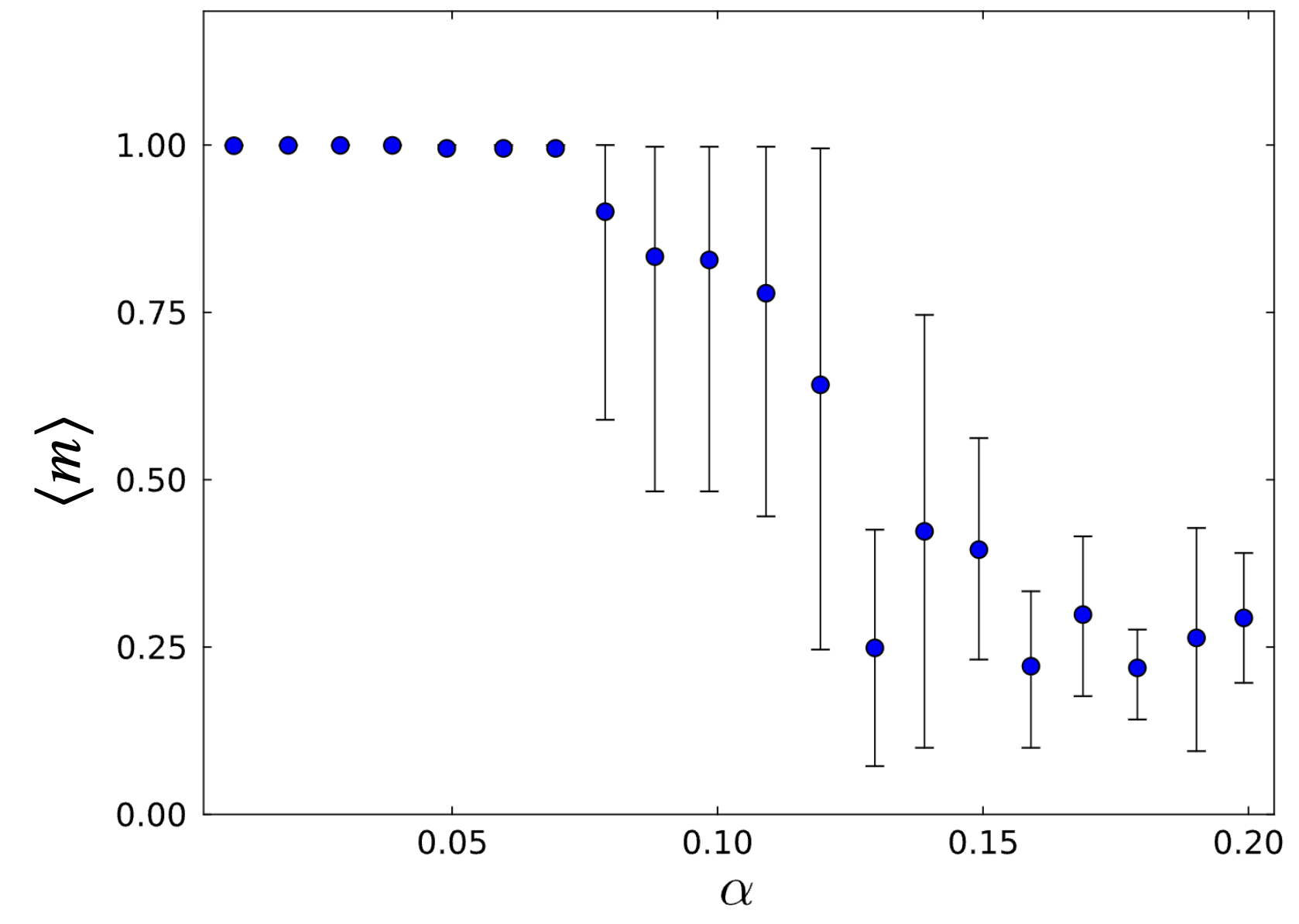}
        \caption{}
    \end{subfigure}
    \begin{subfigure}{0.45\textwidth}
        \centering
        \includegraphics[width=0.99\linewidth]{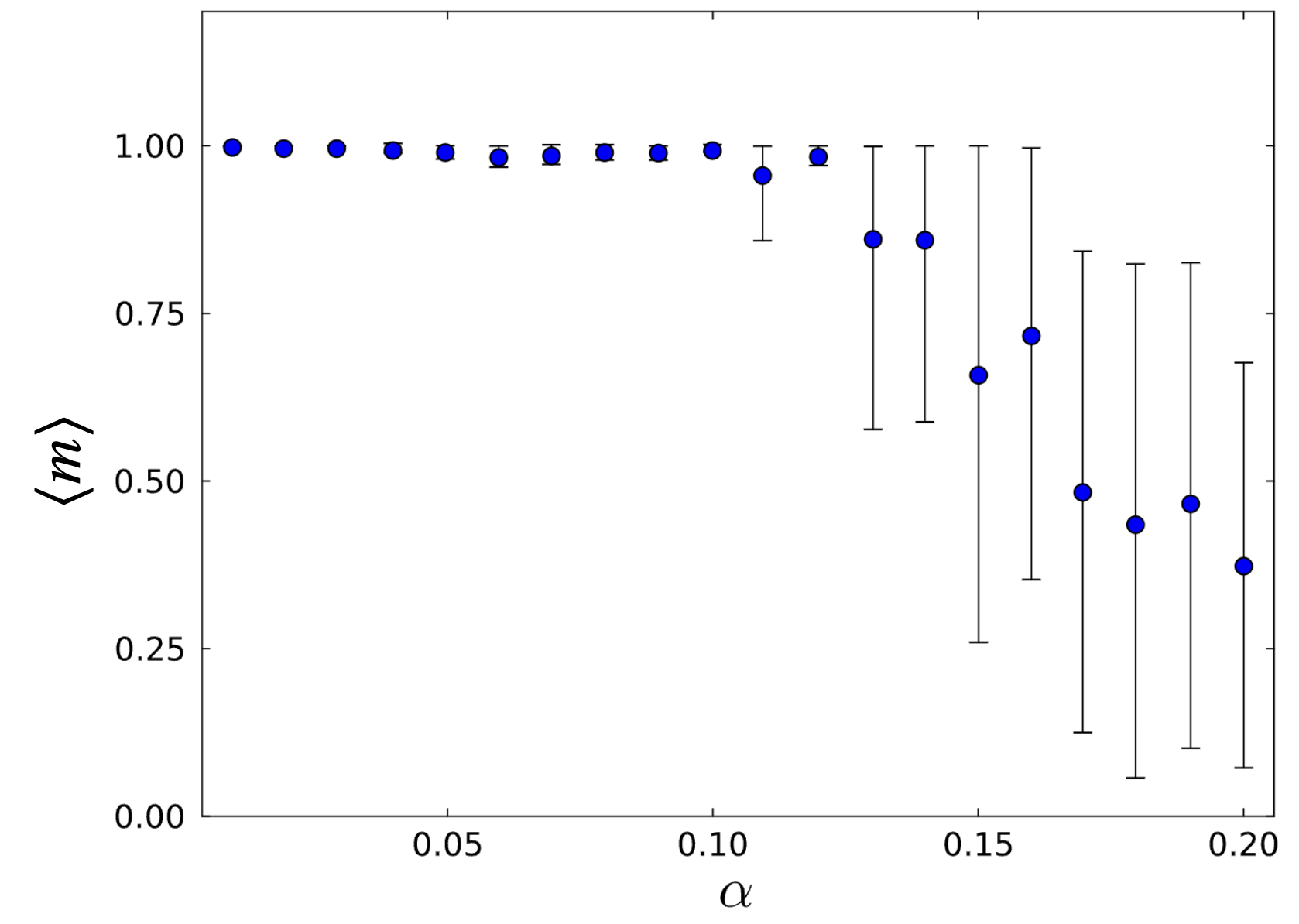}
        \caption{}
    \end{subfigure}
        \caption{The average overlap $m$ plotted against $\alpha$ for different values of $\mathfrak{c}$, where $\epsilon = 0.001, \epsilon_2 = 0.0003$. (a) $\mathfrak{c} = 4$, (b) $\mathfrak{c} = 14$, (c) $\mathfrak{c} = 18$. The error-free retrieval is higher for higher $\mathfrak{c}$. The vertical bars represent maximum and minimum. The network is configured with $N = 200$, and the Rössler parameters are set to $\mathfrak{a} = 0.1$ and $\mathfrak{b} = 0.1$.}
    \label{fig:second_order_coupling_c_4_14_18_20}
\end{figure}
The storage capacity of COAM with $N$ oscillators is numerically investigated by evaluating retrieval performance across a range of loading parameters $\alpha$, using randomly generated patterns.  As before, the retrieval quality is quantified using the average overlap 
$\langle m \rangle$ and its spread.
%
%The storage capacity, characterized by the number of patterns that can be stored with (almost) error-free retrieval in a $N$-oscillator CAM is numerically investigated next. CAM models are generated for various values of $\alpha$ and the retrieval performance is examined for a large set of patterns, all generated randomly.   As in the previous section, the quality of the retrieval is quantified using the average overlap $\langle m \rangle$ and their spread. 
The studies have been carried out for several dynamical regimes of the underlying R\"ossler chaotic oscillators, Fig.\ref{fig:second_order_coupling_c_4_14_18_20} shows the variation of $\langle m \rangle$ as a function of $\alpha$ for three different dynamical regimes: periodic ($\mathfrak{c}=4$), strongly chaotic ($\mathfrak{c}=14$) and extremely chaotic ($\mathfrak{c}=18$). Perfect retrievals are indicated when there is no spread; the storage capacity is taken to be the largest value of $\alpha$ with perfect retrieval. Three key observations emerge. First, for $\mathfrak{c}=4$, which corresponds to a periodic regime of the underlying neurons in the network, the addition of second-order Fourier mode coupling increases the storage capacity to $\alpha=0.04$, matching the capacity of OAMs with $K=2$ and marking a significant improvement over the $K=1$ case. This demonstrates that COAMs generalize OAMs.
Second, as $\mathfrak{c}$ is increased and the system becomes more chaotic, storage capacity improves, indicating enhanced stability of the stored phase-synchronized states—consistent with the trend in $\Re\{\lambda_{\rm max}\}$ shown earlier. Specifically, the storage capacity rises to $0.07$ for $\mathfrak{c}=14$ and to $0.12$ for $\mathfrak{c}=18$. The former case is the reported storage capacity for OAMs
for $K = 3$, while the latter represents a more than $70\%$ increase over the highest reported OAM capacity. Lastly, placing these results in context, the storage capacity of the classical Hopfield network is $\alpha_c =1/(2\ln N)$ \cite{nishikawa}, which when $N=100$ is approximately $0.108$. The observed capacity of $0.12$ in COAM thus exceeds even that of the Hopfield model, highlighting the potential of chaos-enhanced associative memory. The retrieval capacity for the Hopfield network with errors has been theoretically shown to be $0.138$ \cite{hopfieldS}. The storage capacity of $0.12$ for the proposed COAM model with error free retrieval needs to be viewed in this perspective.

\subsection{Biological Significance}
\begin{figure}[htbp]
    \begin{subfigure}{0.45\textwidth}
        \centering
        \includegraphics[width=0.8\linewidth]{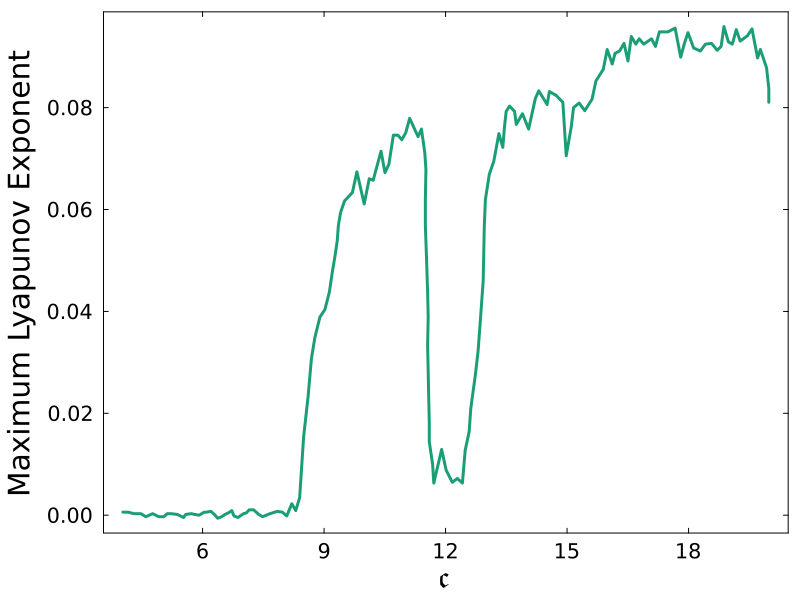}
        \caption{}
        \label{fig:maximum_lyapunov_exponent}
    \end{subfigure}
    \begin{subfigure}{0.45\textwidth}
        \centering
        \includegraphics[width = 0.9\textwidth]{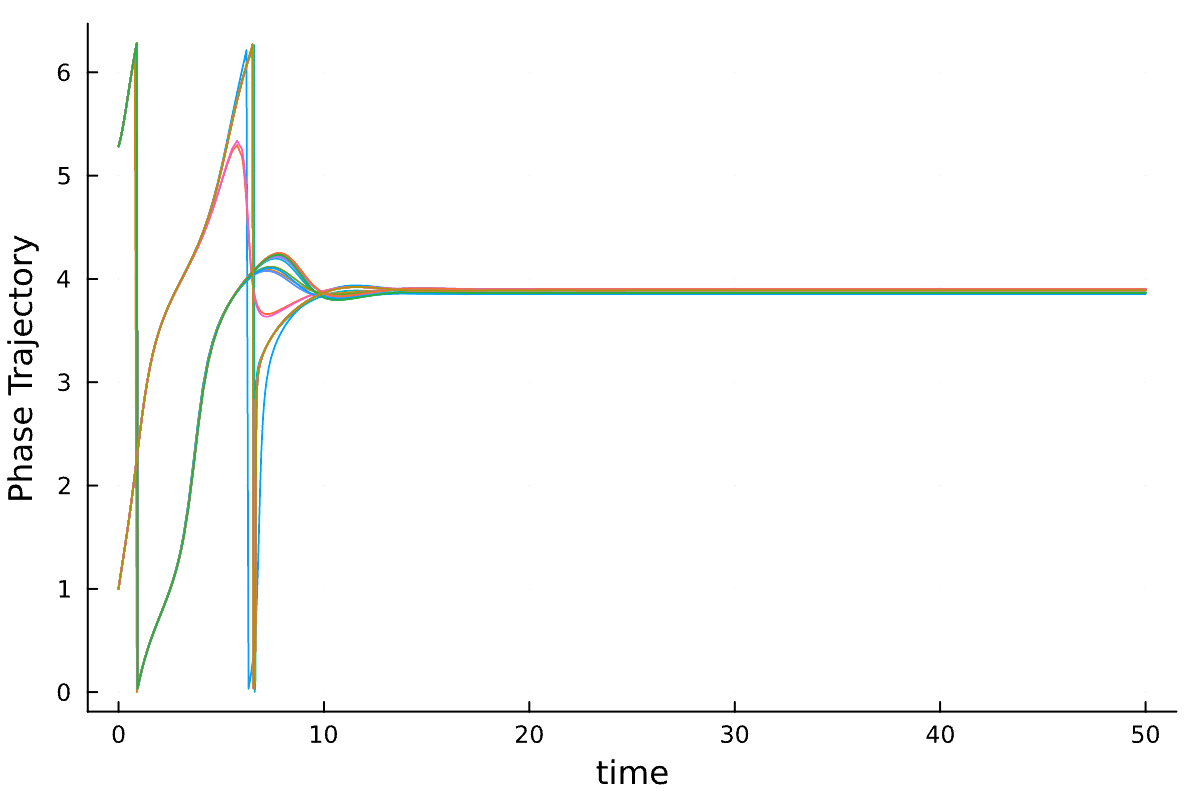}
        \caption{}
\label{fig:Fixed_point_attractor_version}
    \end{subfigure}
    \caption{(a) The maximum Lyapunov exponent is plotted as a function of $\mathfrak{c}$. In the experiments, the network parameters were set to $N = 100$, $p = 6$, $\epsilon = 0.01$, and $\epsilon_2 = 0.003$, while the Rössler parameters were fixed at $\mathfrak{a} = 0.1$ and $\mathfrak{b} = 0.1$.(b) The trajectories of the individual COAM neurons approach a fixed point, for $(\mathfrak{a, f, c}) = (-1.1, 0.2, 5.7)$. The patterns are introduced with an initial overlap $m_0 = 0.6$. As the system is allowed to evolve in time, the trajectories converge to a fixed point representing the deep anesthetic state.}
\end{figure}

The model shows chaotic behavior for $\mathfrak{c}>8.7$. Additionally, the network is capable of showing features similar to the deep anesthetic state as discussed by Skarda and Freeman \cite{skf} for $\mathfrak{a}\leq -1$. The network Lyapunov exponent analysis is performed on the COAM model as a function of $\mathfrak{c}$. The network parameters were set to $N = 100$, $\alpha = 0.06$, $\epsilon = 0.01$ and $\epsilon2 = 0.03$. The R\"ossler parameters, $\mathfrak{a, f}$ were kept fixed at $0.1$. The maximum Lyapunov exponent is plotted as a function $\mathfrak{c}$, see Fig. \ref{fig:maximum_lyapunov_exponent}. It is observed that for $\mathfrak{c} < 8.7$, the system is in a periodic state characterized by the almost $zero$ maximum Lyapunov exponent. But for $\mathfrak{c} > 8.7$, the maximum Lyapunov exponent is positive, indicating that the system is in a chaotic state. Further, for R\"ossler parameters $\mathfrak{a} = -1.1, \mathfrak{b} =0.2, \mathfrak{c} = 5.7$, each of the R\"ossler systems converge to a point attractor, consequently, the COAM model converges to a point attractor. This is similar to the deep anesthetic state discussed in \cite{skf}, see Fig. \ref{fig:Fixed_point_attractor_version}.

In Eq.(\ref{eq:CAM_with_second_order_coupling}), the $\phi$ equation contains the term $\mathfrak{a}\sin 2\phi$ which acts as a subharmonic injection locking term and has been shown to promote antiphase synchronization in networks of coupled oscillators \cite{wang2021solving}. %However, the effect of increasing $\mathfrak{a}$ is not analyzed as in this case, $D_p$ is significantly higher. 
It has been demonstrated that the intrinsic chaotic perturbations in the phase dynamics of the R\"ossler system facilitates a more performative implementation of oscillatory associative memory models. The noise inherent in the R\"ossler system introduced into the phase dynamics $\phi_i$ by the amplitude state variable $A_i$ is comparable to the controlled noise source, which has been suggested to enhance retrieval capability.

\section*{Concluding Remarks}
The chaotic oscillator associative memory model is a generalization of existing oscillator based associative memory models and is inspired from the ideas of Freeman, who first hypothesized that chaotic activity in the brain supports memory retrieval. The model is comprised of a network of chaotic oscillators, whose coupling strengths are based on Hebbian learning of the stored memories. The principle of phase based encoding of memories in OAMs is extended to this model  by developing a phase definition from the higher state chaotic dynamics. Investigations reveal this phase definition to be consistent with the emergent collective dynamical phenomenon of  synchronization based on this defined phase, in the COAM network. The stability of these synchronization states is shown to be strengthened through inclusion of higher order Fourier modes in the coupling. Notably, the numerical simulations presented reveal retrieval capacity of COAM to be significantly higher than existing associative memory models, outperforming even the traditional Hopfield network in terms of perfect memory retrieval. This result is significant as the present  model is suitable for time dependent patterns, that are characteristics of  cognitive tasks. 

The conclusions presented here are general for any COAM, whose individual neurons have low $D_p$ and the couplings are weak. {In networks of chaotic systems with high $D_p$, the presence of multiple characteristic time scales necessitates a substantially higher coupling threshold to achieve phase synchronization. Without this stronger coupling, phase slips may occur as a result of the time-scale disparities between the oscillator's dynamics. Consequently, the onset of chaotic phase synchronization in these high-$D_p$ systems is typically accompanied by the emergence of high-amplitude correlations.}
%n this brief, a Chaotic Neural Associative Memory (CAM) oscillator model based on the chaotic dynamics of the Rössler system is proposed. The model improves its retrieval performance by increasing the parameter $\mathfrak{c}$, and achieves a notable increase in capacity ($\alpha_c = 0.12$) for oscillator-based associative memory models.  The arguments presented in this brief can be extended to chaotic systems with small values of $ D_p $. However, 
In fact, when $D_p$ is high,  phase synchronization is not attainable without achieving generalized complete synchronization, wherein all the state variables of the individual neurons become strongly correlated  and in the limiting case, identical~\cite{Kurths_strong_chaotic_systems}. As phase encoding is not possible in such systems, these are not functional as associative memory models.

\backmatter
\bmhead{Acknowledgements}
Acknowledgments
The first author acknowledges the financial support from  Council for Scientific and Industrial Research towards the PhD program. All the authors acknowledge the partial funding received from the Ministry of Education, Govt of India for funding through the Institute of Eminence scheme with sanction no. 11/9/2019-U.3(A).

\section*{Declarations}
N.R.R., V.S.C., and S.G. conceptualized the problem, performed the research, contributed new reagents/analytic tools,  analyzed data, and wrote the paper. \\

\noindent The authors have no conflicts to disclose. 

% Some journals require declarations to be submitted in a standardised format. Please check the Instructions for Authors of the journal to which you are submitting to see if you need to complete this section. If yes, your manuscript must contain the following sections under the heading `Declarations':

% \begin{itemize}
% \item Conflict of interest/Competing interests (check journal-specific guidelines for which heading to use)
% \item Ethics approval and consent to participate
% \item Consent for publication
% \item Materials availability
% \end{itemize}

% \noindent
% If any of the sections are not relevant to your manuscript, please include the heading and write `Not applicable' for that section. 

% %%===================================================%%
% %% For presentation purpose, we have included        %%
% %% \bigskip command. Please ignore this.             %%
% %%===================================================%%
% \bigskip

% Springer journals and proceedings: \url{https://www.springer.com/gp/editorial-policies}

% Nature Portfolio journals: \url{https://www.nature.com/nature-research/editorial-policies}

% \textit{Scientific Reports}: \url{https://www.nature.com/srep/journal-policies/editorial-policies}

% BMC journals: \url{https://www.biomedcentral.com/getpublished/editorial-policies}
% \end{flushleft}

\begin{appendices}
\section*{Mathematical Description of Oscillatory Associative Memory Models}
\label{Sec:OAMReview}
The mathematical principles of OAM are first discussed as these serve as the foundation for COAM. OAM models are networks of phase oscillators (single state), topologically arranged as a completely connected weighted network. The dynamics of each neuron of a network of $N$ nearly identical phase oscillators is as given in Eq.(\ref{eq:general_oscillator_associative_memory}), and in general, %$\sum_{j=1}^N \mathfrak{F}_{ij}(\theta_i, \theta_j) =$
\begin{align}\label{eq:OAM}
    \sum_{j=1}^N \mathfrak{F}_{ij}(\theta_i, \theta_j)  &=
    \epsilon \sum_{j=1}^N 
    J_{ij} \sin(\theta_j - \theta_i + \Theta_{ij})+ \\
    &\sum_{k = 2}^{K}\left(\frac{\epsilon_k}{N} \sum_{j=1}^N 
    ~\sin~k(\theta_j - \theta_i)\right), 
\end{align}
 $i=1~ \dots ~N. $
The first term on the right hand side represents the pairwise principal Fourier coupling mode, where $\epsilon J_{ij}$ denotes its strength, $\Theta_{ij}$  the phase delay modeling the synaptic delay, the parameter $K \in \mathbb{N}$ represents the number of higher order Fourier modes in the pairwise coupling and $\epsilon_k$  their corresponding strength. The memory information of a pattern is encoded in terms of the phase of the oscillators and manifests as the connection weight, representing synaptic strength. The strength of the primary Fourier mode connections is based on Hebbian learning of $p$ stored patterns, and is given by
\begin{equation}\label{eq:phase_weight_1}
    {\bf W}_{ij} = \frac{1}{N}\sum_{\eta=1}^p \mathfrak{b}^{(\eta)}_i \overline{\mathfrak{b}}^{(\eta)}_j = J_{ij} \exp{(\mathrm{i}\Theta_{ij})}.
\end{equation}
Here,  the $\eta$-th stored pattern  is encoded in terms of the   phase of the oscillators $\{\theta_i^{(\eta)}\}_{i=1} ^N$ and are represented as a complex vector $ \mathfrak{ b}^{(\eta)} = (\mathfrak{b}_1^{(\eta)} \dots \mathfrak{b}^{(\eta)}_N)^\mathrm{T}$,  %each 
where $\mathfrak{b}_i^{(\eta)} = \exp(\mathrm{i}\theta_i^{(\eta)})$, %$\eta = 1 \cdots p$, 
$(\overline{\cdot})$ represents the complex conjugation and i = $\sqrt{-1}$. The loading parameter is defined as $\alpha = p/N$.
The memory retrieval quality is defined in terms of the measure
\begin{equation}\label{eq:continuous_overlap_m_1}
    m = \frac{1}{N}\sum_{i=1}^N \overline{\mathfrak{b}}^{(\eta)}_i s_i.
\end{equation}
that quantifies the overlap between the system state vector given by $s_i = \exp(\mathrm{i}\theta_i)$, $i=1,\hdots, N$ and any pattern $\mathfrak{b}^{(\eta)}$.  %is quantified in terms of an order parameter $m^{(\eta)}$, defined as \begin{equation}\label{eq:continuous_overlap_m}
 %   m^{(\eta)} = \frac{1}{N}\sum_{i=1}^N \overline{\mathfrak{b}}^{(\eta)}_i s_i.
%\end{equation}
In the special case where patterns are binary, {\it i.e.}, %that is, 
$\mathfrak{b^{(\eta)}_i} = \pm 1$ with equal probability,  the system defined by Eq.(\ref{eq:OAM}) has $2^N$ fixed points corresponding to all binary patterns of length $N$.  However, only $2^{N-1}$ are unique as at steady state the network oscillates between a pattern and its negative.  In this form, Eq.(\ref{eq:OAM}) represents Aonishi's model \cite{aonishi} for $K=1$, Nishikawa's model \cite{nishikawa} for $K=2$   and Follman's model \cite{follman} for $K=3$. Given a pattern  $\mathfrak{b}^{(\eta)} = (\mathfrak{b}^{(\eta)}_1 \cdots \mathfrak{b}^{(\eta)}_N )^T$, there is a unique %(up to constant translation) 
fixed point solution that corresponds to the pattern  described in Eq.(\ref{eq:COAM2}).
%\begin{eqnarray}
%    \vert \theta_i - \theta_j \vert  \approx \left\{
%	\begin{array}{ll}
%		0,  & \mbox{if } \mathfrak{b}_i = \mathfrak{b}_j , \\
%		\pi, & \mbox{if } \mathfrak{b}_i \not =  \mathfrak{b}_j .
%	\end{array}
%\right.
%\end{eqnarray}

\section*{Stability Analysis}
\label{SubSec:OAM_Stability}
OAM encodes pattern states through a defined configuration of antiphase synchronization among oscillators. The retrieval of stored patterns is contingent on the stability of the phase-synchronized state. The symmetry of the coupling ensures that the system has a Lyapunov function, which  for the simplest OAM model (for $K=1$) is  
\begin{equation}
\mathcal{L}(\mathbf{\theta; W, \epsilon}) = \mathbf{-} \Re\left(\sum_{i,j = 1}^N W_{i,j} e^{\mathrm{i}\theta_i}e^{-\mathrm{i}\theta_j}\right).
    \label{eq:OAM_single_Lyapunov}
\end{equation}
This can be shown to be a global Lyapunov function. It follows that as $t\rightarrow \infty, ~ \mathcal{\hat{L}}$ approaches a local minimum {\it i.e.}, given any initial condition, the system evolves to a fixed point of Eq.(\ref{eq:OAM}). 

An alternative to using Lyapunov functions to analyze the dynamical stability of the equilibria of these class of systems is through   self-consistent signal-to-noise analysis (SCSNA).  SCSNA is a theoretical framework for evaluating the retrieval quality and capacity of associative models by analyzing the interplay between the stored memory pattern and the distortions caused by noise and/or interference from other patterns. The analysis assumes a large number of stored patterns and considers how each neuron’s activation evolves during retrieval. A self-consistent equation is derived to determine whether a retrieved pattern remains stable under iterative updates and enables estimating the critical storage capacity. A first step involves defining the overlap order parameters $m_c=\frac{1}{N}\sum_i\mathfrak{b}_i^{\eta}\cos\phi_i$ and $m_s =\frac{1}{N}\sum_i\mathfrak{b}_i^{\eta}\sin\phi_i$. Eq.(\ref{eq:general_oscillator_associative_memory}) is  expressed in terms of these order parameters and the equilibrium solution for $\dot{\theta}=0$ is analyzed; see Supplementary Material for more details. The absolute overlap parameter  $m^\eta$ is defined as
\begin{equation}\label{eq:absolute_overlap_parameter}
    m = \left\lvert \dfrac{1}{N} \sum_{j=1}^{N} \mathfrak{b}_j^{(\eta)} e^{\mathrm{i}\phi_j}\right\rvert = \sqrt{m_c^2 + m_s^2}, \,\,\,\eta = 1 \cdots p, 
\end{equation}
and is used as a measure of the overlap of the state of the system with the binary pattern $\mathfrak{b}^{(\eta)}$. For $K=1$ and $\alpha_c=0.42$, the overlap measure $m=0.69$ \cite{aonishi}. The capacity of OAM model can be increased by modifying the coupling function $\mathfrak{F}_{ij}$. Specifically, introducing gaps in the coupling function, such that $\mathfrak{F}_{ij}(\theta_i, \theta_j) =$
\begin{equation}
    \begin{cases} 
    ~\sin (\theta_i - \theta_j),&\theta_i - \theta_j \in (-\frac{\pi}{2} + 2k\pi, \frac{\pi}{2} + 2k\pi),\\
    ~\mu\sin (\theta_i - \theta_j),&\text{otherwise} .
    \end{cases}
    \label{eq:gaps in coupling function}
\end{equation}
improved the storage and retrieval performance, where $\mu$ controls the gaps in the coupling  and $k \in I$. The retrieval performance is enhanced for $\mu = 0.5$ \cite{aonishi}, with $\alpha_c = 0.05$ \cite{aonishi}. 

However, retrieval quality degrades significantly with increasing $\alpha$. While for $ p \leq 2 $ the stored patterns can be retrieved without error, the stored patterns become unstable for higher values of $\alpha$ leading to substantial retrieval errors. This instability arises because, for larger $\alpha$, a minima of Eq.(\ref{fig:lyapunov_function}) is typically located near—but not exactly at—the fixed point corresponding to a stored memory pattern \cite{nishikawa}. The stability of the patterns can be  determined from the sign of the maximum eigenvalue of the Jacobian when Eq.(\ref{eq:OAM}) is linearized at a stored pattern. In particular, $\lambda_{\max}$ remains strictly positive for $ p > 2 $, indicating the loss of stability. Introducing higher-order Fourier modes in the coupling leads to a reduction in the value of $\lambda_{\max}$ thereby improving the stability of a stored pattern.  Specifically, for $ K = 2 $, incorporating second-order coupling with strength $\epsilon_2$ shifts $\lambda_{max}$ by $2\epsilon_2$ towards zero, thereby, improving stability. %It can be shown similarly that 
The stability can be further enhanced by including a third order Fourier mode in the coupling.

\section*{Algorithm for numerical calculations of the average overlap}
\label{appendix-b}
The following algorithm is followed to calculate the average overlap. 
 \begin{enumerate}
    \item The $p$ patterns are constructed as described above, $\mathfrak{b}^{(\eta)} = (\mathfrak{b}_1^{(\eta)} \dots \mathfrak{b}^{(\eta)}_N)^T$, $\eta = 1 \cdots p$. The patterns are stored in the model using Eq.(9).
    \item Select any one of the patterns $(\eta)$ at random, $\eta = 1, \dots, p$.
    \item The state variable $\phi_i$ is initialized,  $\phi_i = \mathfrak{b}_i^{(\eta)}$, in such a way that the overlap is only about 60\%, that is, $m_0^{\eta} = 0.6$.
    \item Subsequently, the governing equations Eq.(14) are evolved in time for long time duration ($10^4$ seconds) and the average overlap over time is calculated at steady state, denoted as $m^{(\eta)'}$.
    \item Repeat steps 3, 4 for 10 different initial conditions to get $(m^{(\eta)'})_{i=1}^{10}$. The average overlap for the chosen pattern is then denoted $m^{(\eta)}$. This procedure is repeated for all $p$ patterns.
    \item Repeat steps (1) - (5) for 100 different sets of patterns, calculate the average overlap $m$, the standard deviation and the maximum and minimum overlap values for $p$ patterns stored. 
    \item Repeat steps (1) - (5) for increasing $p$ to get overlap as a function of number of patterns $m(p)$, or loading rate $m(\alpha)$ .
\end{enumerate} 
This is used as a measure of the performance of our model.

\section*{Implementation Details}
\label{appendix-c}
The COAM model is implemented in Julia \cite{Julia-2017}, using the SciML libraries DifferentialEquations.jl 
 \cite{rackauckas2017differentialequations}, ModelingToolkit.jl \cite{ma2021modelingtoolkit}, and Sundial.jl \cite{gardner2022sundials, hindmarsh2005sundials}.  Runge-Kutta based differential equation solvers were used with a maximum $dt = 10^{-3}$. Algorithm-1 is used to calculate the average overlap, $m$, for all figures in the paper. The Lyapunov exponents are calculated using the ``H2'' algorithm, \cite{lyapunov_spectrum, dsjl}.


% \section{Appendix title 2}




%% For example:

%% \bibitem[Aladro et al.(2015)]{Aladro15} Aladro, R., Martín, S., Riquelme, D., et al. 2015, \aas, 579, A101

%%\end{thebibliography}




\end{appendices}


\bibliography{sn-bibliography}% common bib file

\end{document}